\documentclass{SciPost}

\hypersetup{
    colorlinks,
    linkcolor={red!50!black},
    citecolor={blue!50!black},
    urlcolor={blue!80!black}
}

\usepackage[bitstream-charter]{mathdesign}
\usepackage{tikz}
\usetikzlibrary{arrows.meta}
\usepackage{wasysym}
\usepackage{bm}
\usepackage{graphicx}
\usepackage{xcolor}
\usepackage[normalem]{ulem}
\usepackage[most]{tcolorbox}
\usepackage{needspace}
\usepackage{booktabs}
\usepackage{listings}
\lstdefinestyle{pycode}{%
    language=Python,
    basicstyle=\ttfamily\footnotesize,
    keywordstyle=\color{blue!70!black},
    commentstyle=\color{green!45!black}\itshape,
    stringstyle=\color{red!60!black},
    numberstyle=\tiny\color{gray},
    showstringspaces=false,
    breaklines=true,
    breakatwhitespace=true,
    columns=fullflexible,
    keepspaces=true,
    backgroundcolor=\color{gray!10},
    frame=single,
    framesep=4pt,
    framerule=0.3pt,
    rulecolor=\color{gray!50},
    xleftmargin=6pt,
    xrightmargin=6pt,
    aboveskip=6pt,
    belowskip=6pt,
}
\newcommand{\pyinline}[1]{\begingroup\setlength{\fboxsep}{0pt}%
  \colorbox{gray!15}{\raisebox{0pt}[0.85\ht\strutbox][0.85\dp\strutbox]{%
    \hspace{1pt}\lstinline{#1}\hspace{1pt}}}\endgroup}

\newcommand{\pyinlinebold}[1]{\begingroup\lstset{basicstyle=\ttfamily\bfseries\footnotesize}\lstinline{#1}\endgroup}
\newcommand{\tablefontsize}{\footnotesize}
\newcommand{\figref}[2]{\hyperref[#1]{\autoref*{#1}(#2)}}
\newcommand{\cw}{\circlearrowright}
\newcommand{\ccw}{\circlearrowleft}
\DeclareSymbolFont{usualmathcal}{OMS}{cmsy}{m}{n}
\DeclareSymbolFontAlphabet{\mathcal}{usualmathcal}

\fancypagestyle{SPstyle}{
\fancyhf{}
\lhead{\colorbox{scipostblue}{\bf \color{white} ~SciPost Physics Codebases }}
\rhead{{\bf \color{scipostdeepblue} ~Submission }}

\fancyfoot[C]{\textbf{\thepage}}
}

\newcommand{\nbra}{n_{\rm step}}
\newcommand{\nwalker}{n_{\rm walker}}
\newcommand{\neff}{n_{\rm eff}}
\newcommand{\nsteps}{n_{\rm steps}}
\newcommand{\Mtotal}{\pyinline{M_total}}
\newcommand{\JAX}{\texttt{JAX}}

\providecommand{\ket}[1]{\left|#1\right\rangle}
\providecommand{\bra}[1]{\left\langle#1\right|}
\providecommand{\braket}[2]{\left\langle#1\middle|#2\right\rangle}
\providecommand{\expectation}[1]{\left\langle#1\right\rangle}

\providecommand{\PyGFMC}{\textsc{PyGFMC}}
\providecommand{\repo}{\href{https://github.com/robin-schaefer/PyGFMC}{https://github.com/robin-schaefer/PyGFMC}}

\newtcolorbox{apilistbox}{
  enhanced,
  breakable,
  colback=scipostblue!4!white,
  colframe=scipostdeepblue!70!white,
  opacityframe=0.50,
  interior style={fill=scipostblue!4!white},
  boxrule=0.6pt,
  arc=2mm,
  left=7pt,
  right=7pt,
  top=4pt,
  bottom=4pt,
  before upper={\let\pyinline\pyinlineplain},
  before skip=6pt,
  after skip=6pt,
}

\begin{document}
\pagestyle{SPstyle}

\begin{center}{\Large \textbf{\color{scipostdeepblue}{
\PyGFMC: GPU-accelerated Green's Function Monte Carlo
}}}\end{center}

\begin{center}\textbf{ Robin Sch\"afer
}\end{center}

\begin{center}
Department of Physics, Harvard University, Cambridge, MA, USA
\\[\baselineskip]
\href{mailto:robin_schaefer@fas.harvard.edu}{\small robin\_schaefer@fas.harvard.edu}
\end{center}

\section*{\color{scipostdeepblue}{Abstract}}
\textbf{\boldmath{%
\PyGFMC{} is a \JAX-based implementation of Green's function Monte Carlo for ground states of sign-problem-free quantum lattice models.
On GPUs it reaches nearly three orders of magnitude higher throughput than on CPUs, and is scalable across multiple devices.
This enables ensembles of up to $10^7$ walkers and expressive guiding wavefunctions, including high-dimensional neural quantum states.
\PyGFMC{} can access ground states of systems in which other quantum Monte Carlo methods struggle to equilibrate.
We demonstrate its capabilities on the notoriously difficult problem of quantum spin ice.
The package provides several built-in Hamiltonians and guiding wavefunctions that can be trained using \texttt{NetKet}.
It computes both diagonal and off-diagonal observables.
Documentation and code are available at \repo{}.
}}




\vspace{10pt}
\noindent\rule{\textwidth}{1pt}
\tableofcontents
\noindent\rule{\textwidth}{1pt}
\vspace{10pt}

\section{Introduction}\label{sec:intro}

Quantum Monte Carlo methods~\cite{becca_quantum_2017,sandvik_quantum_1991} hold the promise of efficiently solving sign-problem-free quantum many-body problems.
Where they apply, their statistical approach makes them the method of choice.
They are neither limited to small system sizes, as exact diagonalization~\cite{weinberg_quspin_2017,weinberg_quspin_2019,schaefer_danceQ_2025,DanceQ_code,wietek_xdiag_2026} is, nor to low dimensionality and area-law entanglement, as tensor network methods~\cite{white_density_1992,schollwoeck_density-matrix_2011,tenpy2024} are.
They also avoid the challenging variational optimization on which neural quantum states~\cite{torlai_learning_2016,carleo_solving_2017,nomura_restricted_2017} rely.
However, while efficient in principle, much of the exotic physics encountered in sign-problem-free models, such as quantum spin liquids~\cite{isakov_spin-liquid_2006,isakov_hard-core_2006,banerjee_unusual_2008,isakov_topological_2011,isakov_universal_2012,kato_numerical_2015,huang_extended_2020,niggemann_gapless_2026} or spin glasses~\cite{rieger_zero-temperature_1994,guo_quantum_1994,rieger_griffiths_1996,guo_quantum_1996,king_quantum_2023,bernaschi_the_2024}, can pose severe challenges in practice.
In particular, when nearly degenerate low-energy states are present, reaching low temperatures or the ground state requires long equilibration times that grow rapidly with system size.
Consequently, obtaining clean statistical data may require substantial computational resources.

We present \PyGFMC{}, a Python library for quantum Monte Carlo, aimed at sign-problem-free models that are hard to equilibrate.
A Hamiltonian $H$ is sign-problem-free if
\begin{align}
	H_{x,x^\prime} = \langle x\vert H\vert x^\prime\rangle \leq 0
\end{align}
for all basis states $\vert x\rangle \neq \vert x^\prime\rangle$.
\PyGFMC{} implements the Green's function Monte Carlo~(GFMC) algorithm~\cite{trivedi_ground-state_1990,calandra_numerical_1998,foulkes_quantum_2001}, a stochastic projection onto the ground state (see also Refs.~\cite{booth_fermion_2009,cufar_rimu_2026}).
By performing a walk with population control, GFMC samples the ground state and provides access to diagonal and off-diagonal observables.
Its advantage over other quantum Monte Carlo methods is the ability to include arbitrarily complex guiding wavefunctions, which are either chosen by hand or optimized numerically beforehand~\cite{qin_combination_2020,lin_greens_2022,lin_greens_2024}.
These approximate the ground state and thereby improve the performance substantially.

\PyGFMC{} is written in \JAX~\cite{jax2018github,flax2020github}, allowing parallelized execution on CPUs and GPUs.
GPU acceleration achieves a speedup of roughly \emph{three orders of magnitude} over the CPU baseline and is scalable across multiple GPUs, making walker populations of up to $10^7$ possible.
This hardware-based improvement makes more expressive guiding wavefunctions possible.
The GPU compatibility naturally suits the \texttt{NetKet}~\cite{SciPostPhysCodeb.7,SciPostPhysCodeb.7-r3.4} infrastructure, which provides access to generic variational states and training environments for guiding wavefunctions.
These can be included in \PyGFMC{} through its modular design, which likewise allows users to add their own Hamiltonian classes.
The following Hamiltonian classes are pre-implemented:
\begin{itemize} 
	\item generic spin-$S$ Hamiltonian (including an optimized class for $S=1/2$)
	\item Bose--Hubbard model
	\item Kitaev's toric code model~\cite{kitaev_fault_2003}
	\item Rokhsar--Kivelson model~\cite{rokhsar_superconductivity_1988}
\end{itemize}
The source code, installation instructions, example notebooks, and the documentation are available at \repo{}.

The paper is organized as follows: \autoref{sec:gfmc} provides a brief overview of the GFMC algorithm, including the role of the guiding wavefunction, and the estimation of diagonal and off-diagonal observables.
It also discusses different sources of errors and how to mitigate them by adjusting the control parameters of the algorithm.
\autoref{sec:examples} applies \PyGFMC{} to concrete models. 
First, as a proof of concept, we solve the one-dimensional transverse-field Ising model and discuss the convergence.
Second, we tackle the notoriously difficult, three-dimensional, and frustrated problem of quantum spin ice~\cite{hermele_pyrochlore_2004,benton_seeing_2012,shannon_quantum_2012,shannon_quantum_2021}. 
For small, ferromagnetic transverse couplings, it hosts a $U(1)$ quantum spin liquid.
We study the phase transition from the liquid into the ferromagnet, which is challenging to equilibrate at low and zero temperature for quantum Monte Carlo due to the small transverse exchange parameters~\cite{kato_numerical_2015,huang_extended_2020}.
Furthermore, we study excitations in the effective model describing the spin liquid.
\autoref{sec:package} documents the package layout and walks through the API.
Lastly, we conclude in \autoref{sec:discussion}.

\section{Green's Function Monte Carlo}\label{sec:gfmc}
The GFMC algorithm is a stochastic projection onto the ground state of sign-problem-free Hamiltonians, using a fixed number of walkers~\cite{trivedi_ground-state_1990,sorella_green_1998,calandra_numerical_1998,becca_quantum_2017}.
Unlike most other quantum Monte Carlo methods, such as the stochastic series expansion~\cite{sandvik_quantum_1991,sandvik_stochastic_1999,syljuaasen_quantum_2002}, GFMC is a ground-state method and does not give access to finite-temperature properties.
However, excited states can be computed in different symmetry sectors of the Hamiltonian, which are, for example, defined by the particle number or a topological quantity (see \autoref{sec:ex:ring}).

We give a brief overview of the algorithm, important details within our implementation, sources of errors, and the control knobs needed for good convergence.
For further reading we recommend Refs.~\cite{calandra_numerical_1998,becca_quantum_2017}.

\subsection{Walker propagation}\label{sec:gfmc:algo}
A fixed number of walkers sample the ground-state amplitudes by exploring different configurations in the Hilbert space.
The algorithm can be understood by means of the power method.
We define the nonnegative matrix
\begin{align}
	T = \lambda \mathbb{1} - H.
\end{align}
Let $H\ket{\Psi_0} = E_0\ket{\Psi_0}$ be the ground state of $H$.
For sufficiently large $\lambda$, $\ket{\Psi_0}$ is the eigenstate of $T$ with the largest eigenvalue in magnitude, and can therefore be obtained by applying $T$ a total of $p$ times to a random initial state $\ket{\Psi}$ with $\vert \langle \Psi\vert\Psi_0\rangle\vert>0$:
\begin{align}
	T^p\ket{\Psi} \xrightarrow{\;p\to\infty\;} \ket{\Psi_0}
\end{align}
which converges to the ground state for sufficiently large $p$.

The projection is carried out by a population of $\nwalker$ \textit{walkers} represented in a basis of the Hamiltonian, e.g., the computational $S^z$ basis.
Each application of $T$ acts as a stochastic process whose matrix entries determine the transition probabilities of every walker.
Given a current configuration $\ket{x}$, the probabilities of transitioning to a new configuration $\ket{x'}$ or remaining in the same configuration are
\begin{align}
	p_{x',x} \;=\; -\frac{h_{x',x}}{b_x}\geq 0
	\qquad\text{and}\qquad
	p_{x,x} \;=\; \frac{\lambda - h_{x,x}}{b_x}\geq 0
	\qquad\text{with}\qquad
	b_x \;=\; \lambda - \sum_{x'} h_{x',x},\label{eq:prob}
\end{align}
where $h_{x',x} = \bra{x'} H \ket{x}$ are the matrix elements of the Hamiltonian connecting $\ket{x}$ to $\ket{x'}$.
The sum runs over all $x'$ (including $x' = x$).
We refer to each application of $T$ as a \textit{transition}, whether or not the configuration changes.

The probabilistic interpretation requires the positivity of all entries of $T$, which is guaranteed by the sign-problem-free property of the Hamiltonian: $-h_{x',x} \geq 0$ for $\ket{x}\neq \ket{x'}$.
$\lambda$ must be chosen large enough that $\lambda - h_{x,x} \geq 0$ for every configuration visited, allowing mixed signs in the diagonal part of the Hamiltonian.

Since the transition matrix is not stochastic, correct sampling requires a weight attached to each walker given by the column sum $b_x$.
For each transition out of $\ket{x}$ (whether or not it enters a new configuration), the walker acquires a weight $b_x$ that must be tracked.
Note that if $T$ were a truly stochastic matrix (nonnegative entries with columns summing to one), every walker would acquire the same weight and the weights would not need to be considered.

For large enough $p$, the walker ensemble, weighted by these accumulated factors, samples the amplitudes of the ground-state wavefunction $\ket{\Psi_0}$.
Note that the sign-problem-free structure of $H$ guarantees that the ground-state wavefunction has no relative signs or phases.

\subsection{Branching}\label{sec:gfmc:branching}
Different walkers accumulate different weights over the course of their evolution.
Since each weight is a product over many transitions, the spread of the weights grows exponentially, and a few walkers dominate the ensemble while most others contribute only negligibly.
This leads to a collapse of the effective sample size.
To circumvent this, the population is systematically controlled and walkers that have accumulated large weights are replicated, while those with small weights are removed~\cite{ghanem_population_2021,chen_algorithm_2023}.
The ensemble size is fixed.

The branching step is performed every $\nbra$ transitions, by which point the $i$-th walker has accumulated the weight
\begin{align}
	w_i \;=\; \prod_{k=1}^{\nbra} b_{x_{i,k}},
\end{align}
where $b_{x_{i,k}}$ is the column sum encountered at the $k$-th transition among the walker's last $\nbra$ transitions.
We use a \emph{correlated} branching scheme where a single random number $\zeta \sim U(0,1)$ generates the $\nwalker$ quantiles $z_\alpha = (\alpha + \zeta)/\nwalker$ for $\alpha = 0, \dots, \nwalker - 1$, and the new ensemble is drawn from the cumulative distribution induced by the weights $\{w_i\}$ at these $\{z_\alpha\}$.
This systematic scheme reduces the variance without introducing bias~\cite{calandra_numerical_1998,becca_quantum_2017}.
Notably, when all weights are equal every walker survives the branching step unchanged.

\begin{figure}[t]
	\centering
	\begin{tikzpicture}[
		>=Stealth,
		walker/.style={draw, circle, minimum size=9mm, inner sep=0.5pt},
	]
		\def\xin{-4.4}
		\def\xa{0}
		\def\xb{3.2}
		\def\xout{4.8}
		\def\xform{5.3}
		\node[walker] (a1) at (\xa, 3.0) {$1$};
		\node[walker] (a2) at (\xa, 2.0) {$2$};
		\node[walker] (a3) at (\xa, 1.0) {$3$};
		\node[walker] (a4) at (\xa, 0.0) {$4$};
		\node[above, inner sep=2pt] at (\xa+0.65, 3.0) {$e_{x_1}$};
		\node[above, inner sep=2pt] at (\xa+0.65, 2.0) {$e_{x_2}$};
		\node[above, inner sep=2pt] at (\xa+0.65, 1.0) {$e_{x_3}$};
		\node[above, inner sep=2pt] at (\xa+0.65, 0.0) {$e_{x_4}$};
		\node[walker] (b1) at (\xb, 3.0) {$1$};
		\node[walker] (b2) at (\xb, 2.0) {$1$};
		\node[walker] (b3) at (\xb, 1.0) {$3$};
		\node[walker] (b4) at (\xb, 0.0) {$4$};
		\draw[->] (\xin, 3.0) -- node[above,midway] {$w_1$} (a1.west);
		\draw[->] (\xin, 2.0) -- node[above,midway] {$w_2$} (a2.west);
		\draw[->] (\xin, 1.0) -- node[above,midway] {$w_3$} (a3.west);
		\draw[->] (\xin, 0.0) -- node[above,midway] {$w_4$} (a4.west);
		\draw[->] (a1) -- (b1);
		\draw[->] (a1) -- (b2);
		\draw[->] (a3) -- (b3);
		\draw[->] (a4) -- (b4);
		\draw[->] (b1.east) -- (\xout, 3.0);
		\draw[->] (b2.east) -- (\xout, 2.0);
		\draw[->] (b3.east) -- (\xout, 1.0);
		\draw[->] (b4.east) -- (\xout, 0.0);
		\node[align=center] at ({(\xin+\xa)/2}, 4.25)
			{weights accumulated \\ over $\nbra$ transitions};
		\node at ({(\xa+\xb)/2}, 4.25) {$n$-th branching};
		\node[anchor=west] at (\xform, 2.0)
			{$\bar w_n \;=\; \tfrac{1}{\nwalker}\sum_i w_i$};
		\node[anchor=west] at (\xform, 1.0)
			{$\bar e_n \;=\; \tfrac{1}{\nwalker\,\bar w_n}\sum_i w_i\,e_{x_i}$};
	\end{tikzpicture}
	\caption{Illustration of a branching step on a four-walker ensemble.  
		The	weights $w_i$ above the incoming arrows are accumulated over the
	preceding $\nbra$ transitions. 
		Each walker also carries a local energy $e_{x_i} = \lambda - b_{x_i}$.  
		From these the average weight $\bar w_n$ and weighted local energy $\bar e_n$ of \autoref{eq:avg_weight} are computed and recorded for the energy estimator \autoref{eq:E0}. 
		The correlated reconfiguration then redraws $\nwalker$ walkers from the distribution of weights: walker $1$ has a large weight and is replicated (two outgoing arrows); walker $2$ has a small weight and is dropped (no outgoing arrow).
	}
	\label{fig:branching}
\end{figure}
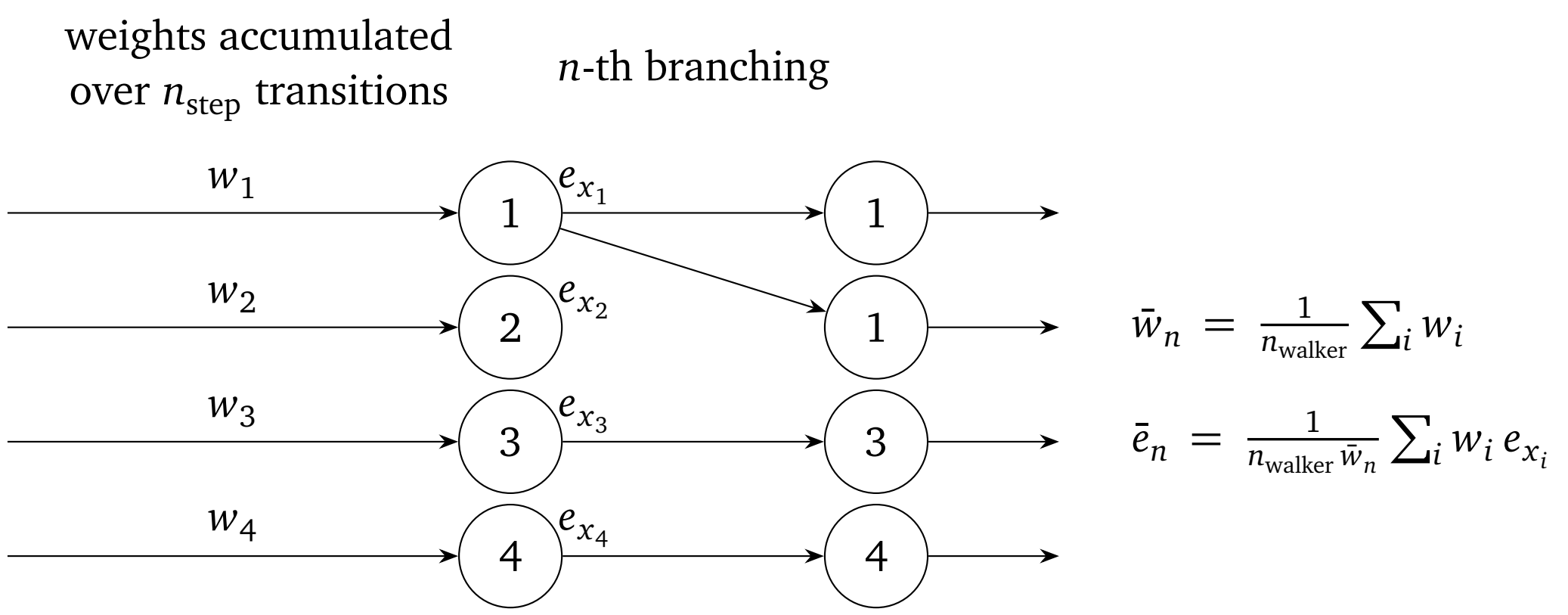

Before the $n$-th branching step we compute the average weight $\bar{w}_n$ across the ensemble and the weighted local-energy estimator $\bar{e}_n$:
\begin{align}
	\bar{w}_n \;=\; \frac{1}{\nwalker} \sum_{i=1}^{\nwalker} w_i,
	\qquad
	\bar{e}_n \;=\; \frac{1}{\nwalker\,\bar{w}_n}
		\sum_{i=1}^{\nwalker} w_i\, e_{x_i}
	\qquad\text{with}\qquad
	e_{x_i} \;=\; \lambda - b_{x_i},\label{eq:avg_weight}
\end{align}
where $b_{x_i}$ is evaluated on the current configuration of the $i$-th walker.
The weights are set to unity after the branching.
\autoref{fig:branching} illustrates the branching step on a four-walker example.

The final ground-state energy estimate as a function of $p$ branching steps is obtained from a single Markov chain by evolving the walker ensemble and recording $(\bar{w}_n, \bar{e}_n)$ at every branching step:
\begin{align}
	E_0[p] \;=\; \frac{\sum_n G_n\, \bar{e}_n}{\sum_n G_n}
	\qquad\text{with}\qquad
	G_n \;=\; \prod_{j=0}^{p-1} \bar{w}_{n-j}.\label{eq:E0}
\end{align}
Note that from here on $p$ counts \emph{branching steps} rather than individual applications of $T$.

\subsection{Propagator}\label{sec:gfmc:cont_time}
\PyGFMC{} offers two schemes for accelerating walker propagation.
One is discrete and based on the single-step walker update described above, whereas the other is formally obtained in the limit $\lambda \to \infty$, leading to a continuous imaginary-time variable.

\paragraph{Discrete formulation.}  
Using the diagonal-stay probability $p_d=p_{x,x}$ from \autoref{eq:prob}, we sample the number $k$ of consecutive diagonal moves before the walker finally transitions to an off-diagonal configuration.
$k$ is geometrically distributed and can be sampled via
\begin{align}
	k \;=\; \left\lfloor \frac{\log(1-\psi)}{\log p_d} \right\rfloor,
	\label{eq:cont_time_discrete_k}
\end{align}
where $\psi \sim U(0,1)$ and $\lfloor \cdots \rfloor$ denotes the floor function.
The walker accumulates the weight $b_x^{\,k+1}$ from $k$ diagonal moves and the subsequent off-diagonal transition, after which it is propagated to a new configuration $\ket{x'}$ drawn from the off-diagonal part of \autoref{eq:prob}.
This procedure repeats until $\nbra$ substeps are reached.

\paragraph{Continuous formulation.} In the limit of $\lambda \to \infty$ with fixed imaginary-time length $\beta = p/\lambda$ the discrete power method turns into the exact imaginary-time propagator $e^{-\beta H}$,
\begin{align}
	(\lambda\mathbb{1} - H)^p
	  \;=\; \lambda^p \left(\mathbb{1} - \tfrac{H}{\lambda}\right)^{\lambda\beta}
	  \;\xrightarrow[\lambda \to \infty]{}\;
	\lambda^p\,e^{-\beta H},
	\label{eq:cont_time_limit}
\end{align}
so that $p$ steps realize an imaginary-time evolution $e^{-\beta H}$ up to a configuration-independent prefactor $\lambda^p$ that cancels in the end.
In this limit, the discrete hold count of \autoref{eq:cont_time_discrete_k} becomes a continuous variable,
\begin{align}
	\mathrm{d}\beta \;=\; \frac{\log(1-\psi)}{\,\sum_{x^\prime\neq x}h_{x^\prime,x}\,},
	\label{eq:cont_time_tau}
\end{align}
where $\sum_{x^\prime\neq x}h_{x^\prime,x} < 0$ is the sum over all off-diagonal elements from $\ket{x}$.
Since $\log(1-\psi) < 0$, the hold time $\mathrm{d}\beta$ is positive.
Over a duration of $\mathrm{d}\beta$ the walker accumulates the weight $e^{-e_{x}\,\mathrm{d}\beta}$, after which it transitions to a new configuration with the same off-diagonal probabilities as in the discrete formulation.
The procedure repeats until the imaginary-time step $\beta$ is reached.
Note that $\lambda$ no longer enters the formalism.\\

\noindent\PyGFMC{} implements three different propagators, which can be chosen using the \pyinline{"method"} keyword:
\begin{itemize}
	\item \pyinline{"single_step"}: the elementary single step.
	\item \pyinline{"multi_step"}: multiple discrete steps at once (discrete formulation).
	\item \pyinline{"cont_time"}: continuous imaginary-time variable (continuous formulation).
\end{itemize}
Neither accelerating scheme (\pyinline{"multi_step"} or \pyinline{"cont_time"}) is GPU-optimal, since both rely on a \pyinline{while}-loop.
However, both still deliver substantial speedups whenever the diagonal-stay probability $p_d$ of \autoref{eq:prob} is non-negligible.

\subsection{Guiding wavefunction}\label{sec:gfmc:guide}
The performance of the algorithm can be significantly improved by providing a guiding wavefunction~$\psi_T$ that approximates the ground-state amplitudes, $\psi_T(x) \approx \braket{x}{\Psi_0}$.
In some rare cases the ansatz can also eliminate a sign problem present in the original basis, for instance via the Marshall sign rule~\cite{marshall_antiferromagnetism_1955}.

The guiding wavefunction alters the transition probabilities of \autoref{eq:prob} to
\begin{align}
	p_{x',x} \;=\; -\frac{h_{x',x}}{b_x}
	\;\;\longrightarrow\;\;
	-\frac{\psi_T(x')}{\psi_T(x)}\,\frac{h_{x',x}}{\tilde{b}_x},
	\qquad
	\tilde{b}_x \;=\; \lambda - \sum_{x'}
		\frac{\psi_T(x')}{\psi_T(x)}\,h_{x',x}.
	\label{eq:prob_guide}
\end{align}
Similarly, the energy estimator and observable expectation values (see below) pick up the same factor $\psi_T(x')/\psi_T(x)$.
Crucially, the algorithm requires only the \emph{ratio} of guide amplitudes between two configurations.
Within \PyGFMC{}, the guiding-wavefunction class returns the log-difference $\log[\psi_T(x')/\psi_T(x)] = \log\psi_T(x') -\log\psi_T(x)$ for every possible transition of the Hamiltonian.

When $\psi_T$ is close to the true ground state, statistical fluctuations are suppressed.
In the limit $\psi_T(x)=\vert\langle x|\Psi_0\rangle\vert$, the fluctuations in the local energy and branching weight vanish entirely.
In this ideal case every walker carries the same weight and the same local-energy estimator, equal to the ground-state energy itself.

\subsection{Observables}\label{sec:gfmc:obs}
The walker ensemble lets us evaluate diagonal and off-diagonal observables in the ground state.
We begin by discussing the forward-walker technique required to compute off-diagonal expectation values.
Diagonal observables can be evaluated within the main Markov chain, reducing computational cost significantly.

\begin{figure}[t]
	\centering
	\begin{tikzpicture}[
		>=Stealth,
		walker/.style={draw, circle, minimum size=9mm, inner sep=0.5pt},
	]
		\def\xa{0}
		\def\xb{3.2}
		\def\xc{6.4}
		\node[walker] (a1) at (\xa, 3.0) {$o_1$};
		\node[walker] (a2) at (\xa, 2.0) {$o_2$};
		\node[walker] (a3) at (\xa, 1.0) {$o_3$};
		\node[walker] (a4) at (\xa, 0.0) {$o_4$};
		\node[walker] (b1) at (\xb, 3.0) {$o_1$};
		\node[walker] (b2) at (\xb, 2.0) {$o_1$};
		\node[walker] (b3) at (\xb, 1.0) {$o_3$};
		\node[walker] (b4) at (\xb, 0.0) {$o_4$};
		\draw[->] (a1) -- (b1);
		\draw[->] (a1) -- (b2);
		\draw[->] (a3) -- (b3);
		\draw[->] (a4) -- (b4);
		\node[walker] (c1) at (\xc, 3.0) {$o_1$};
		\node[walker] (c2) at (\xc, 2.0) {$o_1$};
		\node[walker] (c3) at (\xc, 1.0) {$o_1$};
		\node[walker] (c4) at (\xc, 0.0) {$o_3$};
		\draw[->] (b1) -- (c1);
		\draw[->] (b1) -- (c2);
		\draw[->] (b2) -- (c3);
		\draw[->] (b3) -- (c4);
		\node at (\xa, 3.8) {$n-p/2$};
		\node at (\xb, 3.8) {$n-p/2+1$};
		\node at (\xc, 3.8) {$n$};
		\node[align=left, anchor=west] at (\xc+1.1, 1.5) {%
			$\bar o_{n-p/2}\;=\;\tfrac{3\,o_1 + o_3}{4}$};
	\end{tikzpicture}
	\caption{Forward-walking estimator illustrated for four	walkers over $p/2 = 2$ branching steps of the auxiliary chain.
		The operator is	applied at step $n-p/2$, attaching the estimator $o_x$ to each walker. 
		The estimators are updated following the reconfiguration at every step, where replicated walkers inherit their parent's $o_x$.
		The walker mean of the surviving estimators at step $n$ defines $\bar o_{n-p/2}$ in	\autoref{eq:obs_oavg}.
	}
	\label{fig:fw_ancestry}
\end{figure}
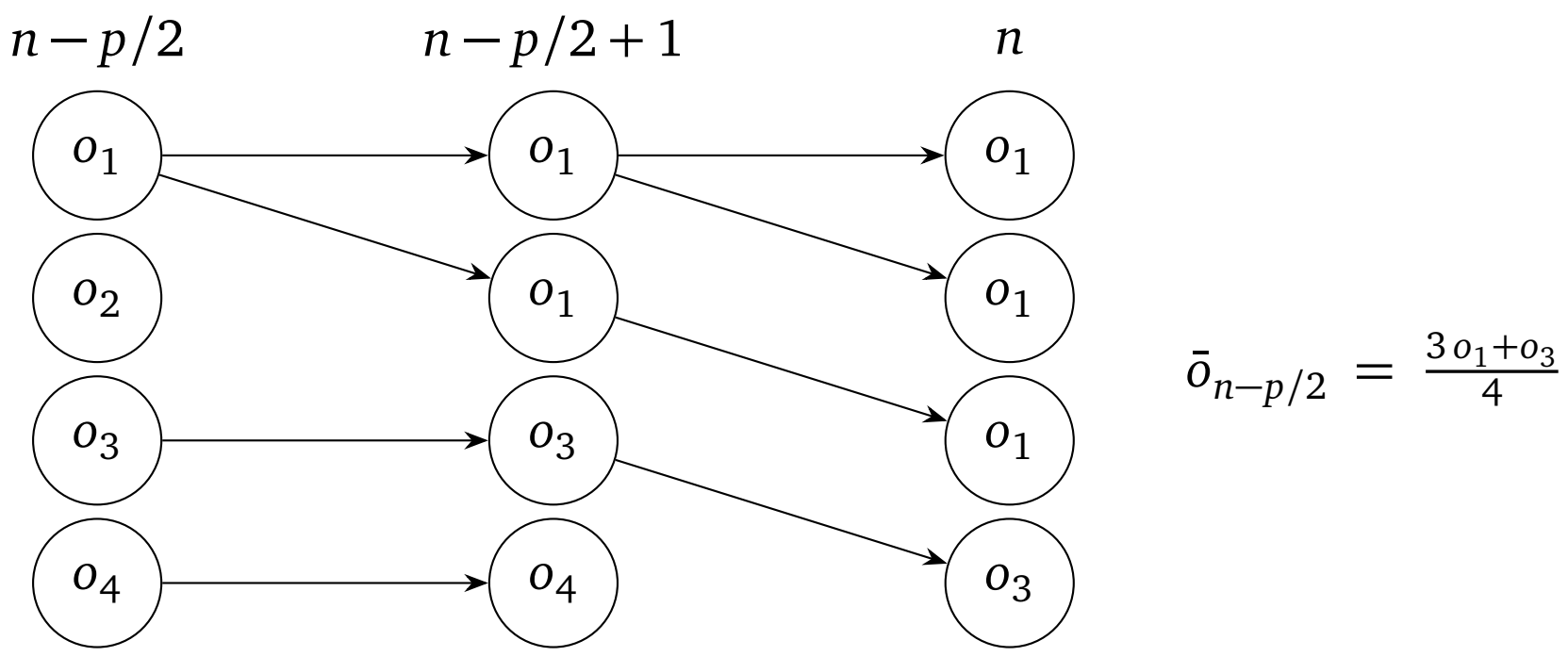

\paragraph{Off-diagonal observables.}
To measure an off-diagonal operator $O$, a fresh short Markov chain of length $p/2$ must be launched.
Let $\ket{x}$ be the current configuration of a walker.
The operator $O$ is applied stochastically, and a target configuration $\ket{x'}$ is drawn with probability
\begin{align}
	p_{x',x} \;=\;\frac{\psi_T(x')}{\psi_T(x)}\,\frac{o_{x',x}}{o_x},
	\qquad
	o_x \;=\; \sum_{x'} \frac{\psi_T(x')}{\psi_T(x)}\,o_{x',x},
	\label{eq:prob_operator}
\end{align}
where $o_{x',x} = \bra{x'} O \ket{x}$ are the matrix elements of $O$.
The configuration of the walker is updated accordingly and the estimator $o_x$ is assigned to it.
The updated ensemble is then evolved under the Hamiltonian for $p/2$ branching steps.
At each branching event a replicated walker duplicates its estimator $o_x$, while a discarded walker's estimator is lost.
This procedure is sketched in \autoref{fig:fw_ancestry}, where four walkers are evolved over two branching events, and the estimator values attached to each walker propagate according to the reconfigurations of the auxiliary chain.
After $p/2$ steps we record the mean of the inherited $o_x$ values
\begin{align}
	\bar{o}_{n-p/2} \;=\; \frac{1}{\nwalker} \sum_{i=1}^{\nwalker} o_{x_i(n)},
	\label{eq:obs_oavg}
\end{align}
together with the branching-weight product of the auxiliary chain
\begin{align}
	G^{\rm aux}_n \;=\; \prod_{j=0}^{p/2-1} \bar w^{\rm aux}_{n-j}
	\;=\; \bar w^{\rm aux}_{n-p/2+1}\,\cdots\,\bar w^{\rm aux}_{n}\,.
	\label{eq:obs_gaux}
\end{align}
The final estimator then resembles \autoref{eq:E0}, but combines $p/2$ weights of the main chain with $p/2$ weights of the auxiliary chain:
\begin{align}
	O[p] \;=\; \frac{\sum_n G^{\rm half}_n \, G^{\rm aux}_n \, \bar{o}_{n-p/2}}{\sum_n G_n}
	\qquad\text{with}\qquad
	G^{\rm half}_n \;=\; \prod_{j=p/2}^{p-1} \bar{w}_{n-j}.
	\label{eq:obs}
\end{align}
$G^{\rm half}_n$ denotes the half of $G_n$ covering the steps $n-p+1,\ldots,n-p/2$.
The other half is obtained from the auxiliary-chain weights $G^{\rm aux}_n$.
The main-chain $G_n$ in the denominator is the same as in \autoref{eq:E0}.

The branching in \autoref{eq:prob_operator} requires nonnegative matrix elements of $O$.
If $O$ has negative matrix elements, it can be decomposed into positive and negative parts, $O = O_{\rm pos} + O_{\rm neg}$, and each contribution is computed separately:
\begin{align}
	\expectation{O} \;=\; \expectation{O_{\rm pos}}	\;-\; \expectation{-O_{\rm neg}}\,.\label{eq:observable_splitting}
\end{align}
The same trick handles complex operators by splitting them into real and imaginary parts.

Note that launching an auxiliary chain to measure the off-diagonal operator increases the runtime by a factor of $p/2$ per observable.

\paragraph{Diagonal observables.}
No auxiliary Markov chain is needed to compute the expectation value of a diagonal operator $O$ since it does not trigger a transition.
We can simply obtain the estimator from the main Markov chain.
At branching step $n-p/2$ we attach to every walker $i$ its current diagonal element $o_{x_i} \equiv \bra{x_i} O \ket{x_i}$ and follow these estimators through the next $p/2$ branching events of the main chain.
The walker mean at step $n$ defines $\bar o_{n-p/2}$ as in \autoref{eq:obs_oavg}, and the diagonal estimator $O[p]$ has the same form as \autoref{eq:obs} except that the auxiliary-chain product $G^{\rm aux}_n$ is replaced by the corresponding main-chain product $\prod_{j=0}^{p/2-1} \bar w_{n-j}$ since both halves of the projection are along the main Markov chain:
\begin{align}
	O[p] \;=\; \frac{\sum_n G_n \, \bar{o}_{n-p/2}}{\sum_n G_n}.
\end{align}

No probabilistic branching of $O$ is involved here, so the diagonal pipeline imposes no positivity restriction on $O$.

\subsection{Convergence and error control}\label{sec:gfmc:simulation}

Obtaining well-converged results is difficult.
It requires understanding why a simulation has failed (or is converging poorly) and how to improve it.
Therefore, we first discuss the two main sources of error and how to identify them, and then explain how to mitigate them.

\subsubsection{Sources of error}\label{sec:errors}
The two sources of error are (i) a systematic bias from an insufficient projection depth and (ii) statistical uncertainty.

\paragraph{(i) Insufficient projection.}
The energy and every observable must be examined as a function of projection steps (branching events).
The estimator initially drifts until it settles to a plateau.
The data have converged when the plateau is stable over several consecutive values of $p$.
However, when the number of projection steps $p$ is pushed too large, the plateau deteriorates, either through a slow drift or rapid changes.
This signals a failure of convergence, and the data cannot be trusted (see below).
Importantly, the deterioration can happen before the plateau is reached, meaning that the simulation is not capable of reliably resolving the ground state with the current settings.

\paragraph{(ii) Statistical errors.}
The deterioration comes from a population collapse in which too many walker lineages have died out, ruining the statistics at deep projection depths.
In fact, after sufficiently many branching events, the entire population descends from a single initial walker.
\PyGFMC{} provides diagnostics for estimating the largest reliably accessible projection depth.
These can be obtained using the \pyinline{statistics} function that analyzes the evolution of a given walker population.
It returns a dictionary \pyinline{stats} (and an evolved walker ensemble).
The two most important entries are:
\begin{itemize}
	\item \pyinline{stats['unique_fraction']}: the fraction of unique configurations present after each branching step.
		For the correlated branching scheme employed here, we recommend that this fraction remain above roughly $0.9$.
		When the step size between two branching events is too large, a few walkers dominate the evolution and many others die out.
		Therefore, the number of distinct configurations decays.
		Conversely, when the step size is too small, the walkers have too little time to accumulate different weights and essentially none die out.
		The fraction stays close to $1$.
	\item \pyinline{stats['effective_sample_size']}: the effective sample size counts how many of the initial walkers effectively contribute descendants after a given number of branching steps.
		We define $c_i$ as the number of present walkers descending from the $i$-th walker.
		The effective sample size is
		\begin{align}
			n_{\rm eff}(p) = \frac{\left(\sum_i c_i\right)^2}{\sum_i c_i^2}= \frac{\nwalker^2}{\sum_i c_i^2}\in [1,\nwalker].\label{eq:neff}
		\end{align}
		If the entire population has collapsed after $p$ steps (with $c_i=\nwalker$ for one $i$ and zero otherwise), $n_{\rm eff}=1$, indicating that $p$ has been pushed too large.
		If, on the other hand, no walker has died out after each step ($c_i=1$ for all $i$), $n_{\rm eff}=\nwalker$ and the statistics are optimal.
		In the single-step example of \autoref{fig:branching}, we have $c_1=2$, $c_2=0$, and $c_3=c_4=1$, leading to $\neff(1)=16/6$.
		We recommend that $n_{\rm eff} \gtrsim 50$ in order to obtain sufficient statistics.
		However, this may vary between different observables and models.
		Note that an observable is measured at the midpoint of its projection window, so the relevant depth is $m=p/2$ rather than the full $p$.

\end{itemize}

The other source of statistical noise is an insufficient number of measurements.
The algorithm is run for $\nsteps$ branching steps, which must be chosen large enough to resolve the numerator and denominator of the estimator to the required precision.
For a well-adjusted simulation, the error bar decreases as $1/\sqrt{\nsteps}$, where $\nsteps$ is the number of measurements (branching steps) performed.
However, since consecutive measurements are correlated, only a reduced, effective number (defined through the autocorrelation time~\cite{madras_pivot_1988}) of measurements is statistically independent.

The autocorrelation time is returned by the production run (see \autoref{sec:pkg:driver}) as part of the \pyinline{stats} dictionary.
It is computed for the logarithmic weights, the energy estimator, and every measured observable.
Given a measurement series $O_n$ of length $\nsteps$, we compute the normalized autocorrelation function
\begin{align}
	\rho(t) = \frac{C(t)}{C(0)}\,,\qquad
	C(t) = \frac{1}{\nsteps-t}\sum_{n} \left(O_n-\bar{O}\right)\left(O_{n+t}-\bar{O}\right)\,,
	\label{eq:acf}
\end{align}
and sum it to a cutoff $M$,
\begin{align}
	A(M) = \frac{1}{2} + \sum_{t=1}^{M}\rho(t)\,.\label{eq:tau_int}
\end{align}
We choose $M$ self-consistently as the smallest value $M$ satisfying $M\geq 5\,A(M)$.
The effective number of measurements is reduced to ${\nsteps}/({2A})$.
The values are stored in \pyinline{stats} under:
\begin{lstlisting}
stats['autocorrelation/energy_estimator']          # scalar
stats['autocorrelation/log_weights']               # scalar
stats['autocorrelation/diag/<name>']               # one entry per p
stats['autocorrelation/off_diag/<name>']           # one entry per p
\end{lstlisting}
The observables are listed per number of branching steps $p$, which corresponds to the averaged estimator after $m=p/2$ branching steps.
Note that this is just the observable estimator and does not include the weight obtained over these $m$ steps.

\PyGFMC{} estimates the statistical error from a single Markov chain using a blocked jackknife~\cite{flyvbjerg_error_1989,efron_jackknife_1982} by calling \pyinline{evaluate_observables} (see \autoref{sec:pkg:driver}).
We divide the entire Markov chain into $n_{\rm blocks}$ contiguous blocks.
For $b=1,\ldots,n_{\rm blocks}$, we compute the estimator by deleting the $b$-th block from the Markov chain.
We define this reduced estimator as $\Theta_b$.
The jackknife error is obtained from the variance of these estimates,
\begin{align}
	\sigma^2 = \frac{n_{\rm blocks}-1}{n_{\rm blocks}}\sum_{b=1}^{n_{\rm blocks}} \left(\Theta_b-\bar{\Theta}\right)^2\,,
	\qquad \bar{\Theta} = \frac{1}{n_{\rm blocks}}\sum_b \Theta_b\,.
	\label{eq:jackknife}
\end{align}
However, we recommend performing several independent simulations and taking the standard error of their mean to obtain error bars on an observable or on the ground-state energy.

\subsubsection{Control parameters}\label{sec:gfmc:params}
The performance of a GFMC calculation can be greatly improved by carefully adjusting the simulation parameters.
We discuss these control parameters and how to tune them, but first emphasize an important point:
The quality of a simulation, especially for systems that are difficult to equilibrate, depends crucially on the guiding wavefunction.
A good variational ansatz is the key to tackling challenging problems, and effort should be invested in optimizing it before undertaking large-scale simulations.

\paragraph{Projection depth.}
The projection depth $\tau$ is determined by the product of the branching interval ($\nbra$ in the discrete method or $\beta$ in the continuous method) and the number of branching steps $p$ divided by $2$:
\begin{align}
	\tau = \frac{p}{2}\nbra \quad \text{(discrete)}, \qquad \tau = \frac{p}{2}\beta \quad \text{(continuous)}.
\end{align}
Note that $\tau$ in the discrete and continuous methods should not be compared directly.

The goal is to reach the largest possible $\tau$ while maintaining good statistics.
The statistics at a given $\tau$ depend on the step size between branching steps, which must balance two competing demands:
First, if it is too small, walkers do not have sufficient time to accumulate weight, and branching cannot efficiently remove low-weight walkers.
Second, if it is too large, weight accumulates excessively and only a few walkers survive each reconfiguration.

To identify a suitable step size in advance, we recommend scanning a range of step sizes and maximizing the effective sample size $\neff(\tau)$ from \autoref{eq:neff} as a function of $\tau$.
We perform this optimization in \figref{fig:tfim_critical}{a}, which shows $\neff$ as a two-dimensional heat map on a log scale as a function of step size $\beta$ (continuous propagation) and projection depth $\tau$.
It includes contour lines to guide the eye.
We find that, for a given $\tau$, there is an optimum around $\beta\approx 0.15$, and $\neff$ decreases as $\beta$ increases from approximately $0.25$ to $1$.
Note that the computational cost of propagation is usually lower for larger values of $\beta$, and the aim is typically to pick the largest value of $\beta$ with sufficiently large $\neff$.

\begin{figure}[t]
	\centering
	\includegraphics[width=0.7\textwidth]{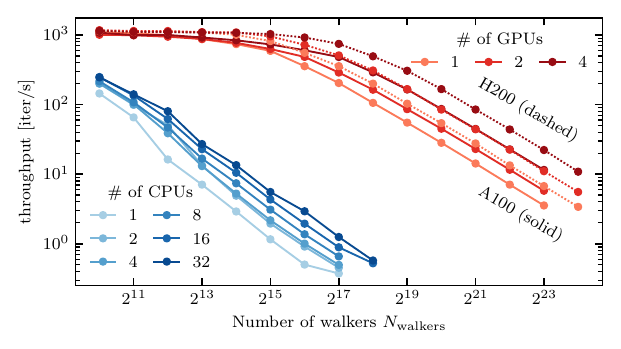}
	\caption{Throughput in iterations per second as a function of the walker population $\nwalker$ for the one-dimensional Heisenberg chain at $L=100$, $\nbra=6$, and $\lambda=25$.  
		Blue points: CPU runs on an Intel Xeon Platinum 8480 using $1$ to $32$ cores.
		Red points: GPU runs on $1$, $2$, and $4$ NVIDIA A100 (solid) and H200 (dashed) devices.  
		Each point is averaged over $10$ independent runs.
	}
	\label{fig:throughput}
\end{figure}

\paragraph{Number of walkers $\nwalker$.}
The number of walkers is the simplest control parameter: a larger population generally yields better statistics and allows larger projection depths.
On modern GPUs, \PyGFMC{} can reach up to $\nwalker \sim 10^7$, which is demonstrated in \autoref{fig:throughput}.
However, blindly scaling up the population is not the best strategy, as it is often more effective to invest in a better guiding wavefunction.
Also note that beyond $10^{7}$, the limited resolution of \pyinline{float32} starts to affect the cumulative weight distribution used in the branching step.

\paragraph{Diagonal shift $\lambda$.}
The diagonal shift $\lambda$ is only present in the discrete methods, and it must be at least as large as the largest diagonal matrix element encountered to ensure positivity.
A larger $\lambda$ increases the diagonal-stay probability $p_d$, so that walkers require many transitions before changing their configuration.

The shift can be set to a value smaller than the maximal diagonal element of the Hamiltonian as long as the configurations with a larger value are practically never visited.
Take a ferromagnet as an example, where the antiferromagnetic N\'eel configuration is highly unlikely and has an exponentially suppressed amplitude.
Hence, it is practically never visited within a finite simulation, and the $\lambda$ value can be safely reduced.

\paragraph{Guiding wavefunction.}
The key ingredient for tackling problems that are hard to equilibrate is a good ansatz wavefunction~\cite{pilati_self-learning_2019,qin_combination_2020,lin_greens_2022,lin_greens_2024}.
The following criteria should be considered:
\begin{itemize}
	\item The ansatz should be close to the true ground state (the overlap should be maximized).
	A simple benchmark for this is a minimal variational energy.
	\item It should not strongly suppress configurations that are required to access different regions of Hilbert space.
	\item The ratio between amplitudes that are connected by the Hamiltonian should not be too large, as this leads to large fluctuations in the simulation.
	\item Symmetries of the Hamiltonian should be respected.
	\item The amplitudes of the wavefunction are evaluated at every possible transition, so they must be cheap to compute.
		Since the algorithm only requires the ratio between the amplitudes of two states connected by a few-body Hamiltonian, we recommend Jastrow-type wavefunctions~\cite{jastrow_many-body_1955,changlani_approximating_2009,mezzacapo_ground-state_2009} of the form:
		\begin{align}
			\log \psi_T(x) =  \sum_{c\in\mathcal{C}} \Phi_c(c[x]),\label{eq:cluster_guide}
		\end{align}
		where the sum runs over a collection $\mathcal{C}$ of clusters $c$.
		$c[x]$ refers to the configuration of $\ket{x}$ on this cluster, and $\Phi_c$ maps a cluster configuration to a real number.

Each Hamiltonian term typically acts on only a small number of sites, so the logarithmic difference between two configurations $\ket{x}$ and $\ket{x^\prime}$ involves only the subset $\mathcal{C}^\prime \subset \mathcal{C}$ of clusters that contain the updated sites:
		\begin{align}
			\log \psi_T(x) - \log \psi_T(x^\prime) =  \sum_{c\in\mathcal{C}^\prime} \left[ \Phi_c(c[x]) - \Phi_c(c[x^\prime]) \right].
		\end{align}
		The bulk of the clusters in $\mathcal{C}$ remains unchanged, which drastically reduces the computational cost.

		The Jastrow-type ansatz further allows for an easy implementation of symmetries, such as translations, by simply setting $\Phi_c=\Phi$ for all $c$.
\end{itemize}

\section{Examples}
\label{sec:examples}
We demonstrate the capabilities of \PyGFMC{} with three examples.
First, we study the one-dimensional transverse-field Ising model to benchmark the convergence of observables~(\autoref{sec:ex:tfim}).
Second, we resolve the transition between a $U(1)$ spin liquid and a ferromagnet on the three-dimensional pyrochlore lattice.
This problem is notoriously difficult to equilibrate due to its small off-diagonal matrix elements~(\autoref{sec:ex:pyro_nn}).
Third, we compute string excitations in the $U(1)$ quantum spin liquid phase on the pyrochlore lattice~(\autoref{sec:ex:ring}).

\subsection{Transverse-field Ising model}
\label{sec:ex:tfim}
We study the one-dimensional transverse-field Ising model on a periodic chain with $N=100$ sites:
\begin{align}
	H_{\rm TFIM} = -J_z\sum_i \sigma^z_i\sigma^z_{i+1} - h_x \sum_i \sigma^x_i\,,
	\label{eq:H_TFIM}
\end{align}
with $J_z=1$.
The model has a zero-temperature phase transition at the critical point $h_x=J_z$, separating a ferromagnetic ($h_x < J_z$) from a paramagnetic ($h_x > J_z$) phase.

We use a single-parameter, nearest-neighbor Jastrow ansatz as a guiding wavefunction:
\begin{align}
	\log\psi_T(x) = -\beta_{\rm Jastrow}(h_x)\sum_i \sigma^z_i\sigma^z_{i+1}\,.\label{eq:jastrow_TFIM}
\end{align}
We cap its magnitude at $|\beta_{\rm Jastrow}|\leq 0.3$ to prevent the walker population from freezing.
This affects the values in the ferromagnetic phase for $h_x \leq 0.9$.
In the paramagnetic phase, $\beta_{\rm Jastrow}=0$ (or equivalently, no guiding wavefunction) is a good ansatz, as it becomes exact for $h_x/J_z\rightarrow \infty$.
The model is exactly solvable and serves as our benchmark~\cite{pfeuty_one-dimensional_1970}.

\begin{figure}[tp]
	\centering
	\includegraphics[width=\textwidth]{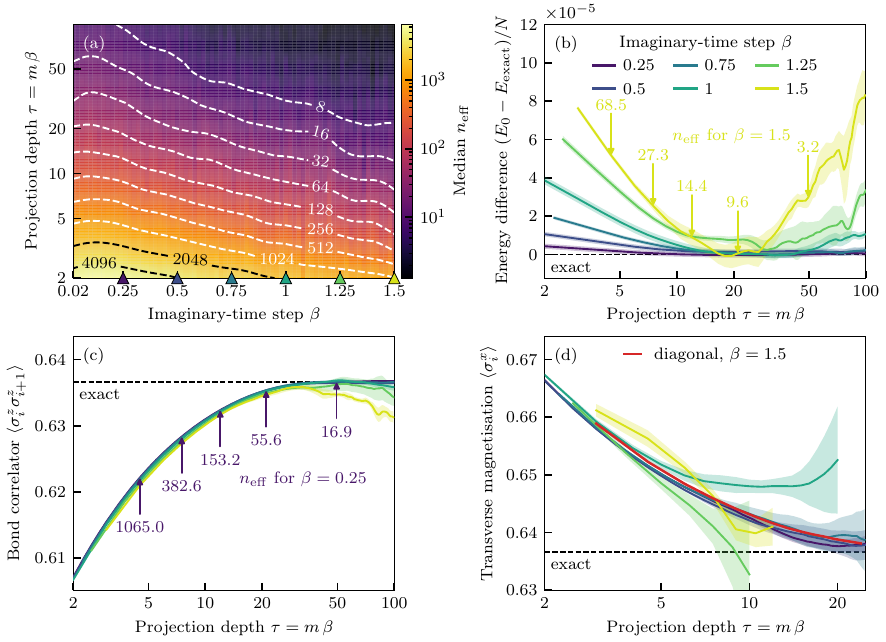}
	\caption{Data for the transverse-field Ising model in \autoref{eq:H_TFIM} at the critical point with $N=100$ on a periodic chain. 
		Panel (a) shows the effective sample size from \autoref{eq:neff} (median over 16 independent runs) on a logarithmic scale as a function of step size $\beta$ (continuous propagation) and the total projection depth $\tau=m\beta$.
		Panel (b) shows the difference from the exact ground-state energy for six different step sizes as a function of total projection depth.
		It further indicates $\neff$ at different values of $\tau$ for $\beta=1.5$.
		Panel (c) shows the averaged nearest-neighbor $(1/N)\sum_i\sigma^z_i\sigma^z_{i+1}$ correlator and marks $\neff$ for $\beta=0.25$.
		The data in panels (b) and (c) were obtained using $\nwalker=2^{17}$ and $\nsteps=2\cdot 10^5$.
		Panel (d) shows the transverse magnetization $(1/N)\sum_i\sigma^x_i$ per site, which has been computed using the off-diagonal forward-walking technique.
		These data were obtained with only $\nwalker=2^{15}$ and $\nsteps=64\cdot 10^4$, and an off-diagonal stride of $64$ ($10^4$ measurements in total).
		The red line shows the magnetization computed from the energy and the correlator using the data from panels (b) and (c) for $\beta=1.5$.
		Error bars show the standard error of the mean over five independent runs.
	}
	\label{fig:tfim_critical}
\end{figure}

\autoref{fig:tfim_critical} presents results at the critical point and examines the optimal choice of step size $\beta$ for continuous-time propagation.
Panel (a) shows the effective sample size $\neff$ from \autoref{eq:neff} (median over 16 simulations) as a function of the step size $\beta$ (x-axis) and projection depth $\tau=m\beta$ (y-axis) on a logarithmic scale.
We find that survival is optimal ($\neff$ is maximal) near $\beta\approx 0.15$ for the tested values of $\tau$.
The effective sample size decreases as $\beta$ increases from approximately $0.25$ to $1$.

Panel (b) shows the convergence of the energy as a function of projection depth for the step sizes marked on the axis of panel (a).
Remarkably, the energy is resolved to a precision of $10^{-5}$ after only a few $\tau$.
The two largest step sizes ($\beta=1.5,1.25$) fail to resolve the plateau to this precision, since the walker distribution collapses beyond $\tau\approx 10$.
For $\beta=1.5$, we mark the $\neff$ values at selected projection depths $\tau$.
The simulations with $\beta=1$ and $\beta=0.75$ lose convergence beyond $\tau\approx 30$ and $\tau\approx 80$, respectively.
The smallest values of $\beta$ ($\beta=0.25,0.5$) resolve the plateau up to the largest projection depth considered, $\tau=100$.

Panel (c) shows the averaged nearest-neighbor correlator $\sigma^z_i\sigma^z_{i+1}$.
The correlator is less precise than the energy, and its initial deviation from the exact value at small $\tau$ is much larger.
All step sizes follow the same curve until the larger step sizes lose convergence.
We mark the effective sample size for $\beta=0.25$, allowing comparison with the $\beta=1.5$ results in panel (b).

Panel (d) shows the off-diagonal transverse magnetization per site.
These data are obtained for fewer walkers and shorter projection depths because the off-diagonal measurement is substantially more expensive, scaling as $\max(\tau)$.
Consequently, convergence is lost earlier than in the diagonal calculations in panels (b) and (c).
However, the transverse magnetization can also be computed from the energy, panel (b), and the $\sigma^z\sigma^z$ correlator, panel (c), and is shown in red for the largest step size $\beta=1.5$.
This shows that a larger walker ensemble is capable of resolving the transverse magnetization more reliably.

\begin{figure}[t]
	\centering
	\includegraphics[width=\textwidth]{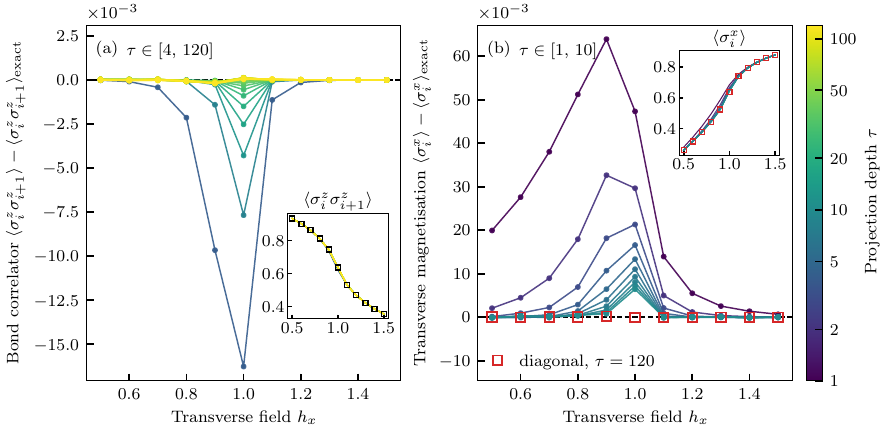}
	\caption{Deviation of the averaged nearest-neighbor correlator $\sum_i\langle\sigma^z_i\sigma^z_{i+1}\rangle/N$ (left) and of the per-site transverse magnetization $\sum_i\langle\sigma^x_i\rangle/N$ (right) from the exact values versus $h_x$ in the transverse-field Ising model from \autoref{eq:H_TFIM} with $N=100$ on a periodic chain. 
	Different colors encode the projection depth $\tau$. 
	The maximal values for the diagonal and off-diagonal observables are $\tau=120$ and $\tau =10$, respectively.
	The red squares show the transverse magnetization obtained from the diagonal correlator and energy at $\tau=120$.
	The same GFMC settings as in \autoref{fig:tfim_critical} were used.
	}
	\label{fig:tfim_zz_xtot}
\end{figure}

\autoref{fig:tfim_zz_xtot} shows the averaged nearest-neighbor correlator and the total transverse magnetization per site at increasing projection depths $\tau$ for different $h_x$ values.
Away from the critical point ($h_x < 0.9$ or $h_x > 1.1$), both observables converge rapidly for $\tau \leq 10$.
At and near the critical point, comparable convergence requires larger projection depths, $\tau \sim 50$.
Our simulations resolve the diagonal correlator reliably across the entire parameter space.
The off-diagonal forward walking was carried out only up to $\tau=10$ and therefore does not converge near the critical point.
However, the transverse magnetization obtained from the diagonal run at $\tau=120$ via $\sum_i\langle\sigma^x_i\rangle = -\left(E_0 +  J_z\sum_i\langle\sigma^z_i\sigma^z_{i+1}\rangle\right)/h_x$ (red squares in \autoref{fig:tfim_zz_xtot}) converges across the entire parameter range considered.

\subsection{Nearest-neighbor XXZ on the pyrochlore lattice}
\label{sec:ex:pyro_nn}
\begin{figure}[t]
	\centering
	 \includegraphics[width=0.7\textwidth]{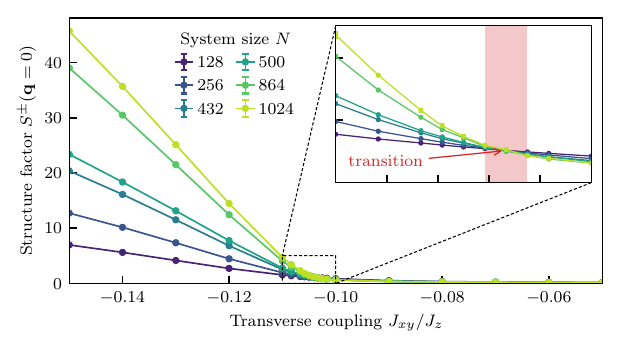}
	\caption{Transverse structure factor at the ordering wavevector, $S^{\pm}(\bm{q}=0)$ from \autoref{eq:Spm_q0}, of the nearest-neighbor pyrochlore XXZ model in \autoref{eq:H_pyro_xxz} as a function of $J_{xy}/J_z$, for different system sizes.
		In the ferromagnetic regime $S^{\pm}(\bm{q}=0)$ grows extensively with $N$, whereas it remains intensive in the spin-liquid phase.
		The inset zooms into the transition region $J_{xy}/J_z\in[-0.11,-0.10]$.
		The shaded band marks the transition at $J_{xy}/J_z=-\frac{31}{300}\pm\frac{1}{1200}$, where the curves for different $N$ cross.
		GFMC parameters: $2^{14}$ walkers ($2^{15}$ for $N=1024$ near the transition), $2000$ measurements with an off-diagonal stride of $\tau\approx 3$ to $\tau\approx 15$, and $2$ independent samples per point.
		We used projection depths from $\tau=14$ to $\tau=60$ depending on $N$ and $J_{xy}$.
		See \autoref{app:pyro_tau} for further information.
	}
	\label{fig:pyro_Spm}
\end{figure}
The pyrochlore lattice is the archetypal three-dimensional frustrated lattice.
It hosts a $U(1)$ quantum spin liquid phase known as quantum spin ice~\cite{hermele_pyrochlore_2004,benton_seeing_2012,shannon_quantum_2012,shannon_quantum_2021}.
The lattice consists of corner-sharing tetrahedra whose centers form a diamond lattice.
We use spin-$1/2$ degrees of freedom coupled by an XXZ Hamiltonian:
\begin{align}
	H = \sum_{\langle i,j\rangle}\Big[\,J_{xy}\,(S_i^x S_j^x + S_i^y S_j^y)
	    \;+\; J_z\, S_i^z S_j^z\,\Big]\,.
	\label{eq:H_pyro_xxz}
\end{align}
It is sign-problem-free for $J_{xy} < 0$.

In the Ising limit $J_{xy}=0$, the model has an extensive classical ground-state degeneracy equivalent to water ice~\cite{pauling_structure_1935,giauque_entropy_1936,anderson_ordering_1956}.
Weak transverse coupling $|J_{xy}|\ll J_z$ lifts this degeneracy through quantum fluctuations described by a ring-exchange Hamiltonian, which arises at third order in perturbation theory~\cite{hermele_pyrochlore_2004}:
\begin{align}
	H_{\rm eff} \;=\; -g\sum_{(i,j,k,l,m,n)\,\in\,\hexagon}
	S_i^+ S_j^- S_k^+ S_l^- S_m^+ S_n^- \;+\;\text{h.c.}\,.
	\label{eq:H_ring}
\end{align}
The sum runs over the hexagonal plaquettes of the pyrochlore lattice and $g\sim |J_{xy}|^3/J_z^2$.
The ring model gives rise to an emergent quantum electrodynamics with photons, electric charges (spinons), and magnetic monopoles.

The system undergoes a phase transition into a ferromagnetic ground state as $|J_{xy}|/J_z$ increases, which is understood as the condensation of spinons.
The transition occurs at a small value, $|J_{xy}|/J_z\approx 0.1036$~\cite{banerjee_unusual_2008}, which makes equilibration difficult and Monte Carlo simulations challenging~\cite{kato_numerical_2015,huang_extended_2020}.
To overcome this difficulty, we employ a Jastrow-type guiding wavefunction that provides a good approximation of the ground state.
We parameterize the guide as $\log\psi_T(x) = \sum_c \Phi(c[x])$, where $c$ runs over all $18$-site clusters formed by the closed loop of six tetrahedra surrounding a hexagonal plaquette.
We refer to these clusters as \emph{crowns}.
Here, $c[x]$ denotes the configuration of the $18$ spins on cluster $c$.
A pyrochlore lattice with $N$ sites hosts $N$ crowns, and each $\Phi_c$ maps the $2^{18}$ configurations to real numbers.
We set the same $\Phi=\Phi_c$ for each cluster and employ various spatial symmetries of the crown cluster that, together with time-reversal symmetry, drastically reduce the number of independent parameters.
Instead of $2^{18}$ coefficients, we end up with only $2110$ independent parameters that must be optimized.
Optimization was performed using NetKet~\cite{SciPostPhysCodeb.7,SciPostPhysCodeb.7-r3.4}.

To resolve the transition into the ferromagnet, we measure the transverse structure factor at $\bm{q}=0$:
\begin{align}
	S^{\pm}(\bm{q}=0) \;=\; \frac{1}{N}\sum_{i<j}\langle S^+_i S^-_j + S^-_i S^+_j\rangle\,.
	\label{eq:Spm_q0}
\end{align}
$S^{\pm}(\bm{q}=0)$ is extensive in the ferromagnetic phase and intensive in the spin liquid, such that the order parameter $S^{\pm}(\bm{q}=0)/N$ remains finite in the ferromagnet and vanishes in the spin liquid as $N\rightarrow \infty$.
\autoref{fig:pyro_Spm} shows the structure factor as a function of $J_{xy}$ for different system sizes.
The inset zooms into the transition region $J_{xy}/J_z\in[-0.11,-0.10]$ and highlights the transition at
\begin{align}
	\frac{J_{xy}}{J_z}=-\frac{31}{300}\pm\frac{1}{1200}\,,
	\label{eq:transition}
\end{align}
which is in agreement with Ref.~\cite{banerjee_unusual_2008}.
\autoref{app:pyro_tau} discusses the convergence with the projection depth $\tau$ and compares the results to the order parameter evaluated for the guiding wavefunction alone (without GFMC).
While the guide captures the overall trend, it fails to resolve the transition reliably.
This illustrates how the GFMC projection and a good variational state complement each other.

\subsection{Ring-exchange on the pyrochlore lattice}
\label{sec:ex:ring}
\begin{figure}[t]
	\centering
	\includegraphics[width=0.7\textwidth]{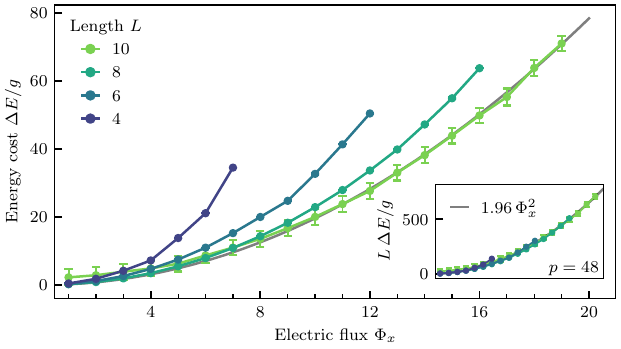}
	\caption{Energy cost of a flux sector $(\phi_x, 0, 0)$ in the pyrochlore ring-exchange model for several linear system sizes $L$ with $N=4L^3$ spins in total.
		The inset shows the $1/L$ scaling collapse.
		GFMC parameters: step size $\nbra=15$, $\lambda=0$, $2^{12}$ walkers and $10^{5}$ steps per flux sector, number of branching steps $p=48$; no guiding wavefunction has been used.
	}
	\label{fig:ring_energy}
\end{figure}

We demonstrate how \PyGFMC{} can access non-trivial excitations by computing the flux cost of the effective ring model of \autoref{eq:H_ring}.
The ring Hamiltonian conserves the total electric flux threading each of the periodic boundaries of the cluster.
Preparing the initial walker population in a chosen flux sector therefore lets us probe the energy cost of flux excitations.
The energy cost of a flux configuration $\bm{\phi}=(\phi_x,\phi_y,\phi_z)$ in a cluster of linear size $L$ scales as $|\bm{\phi}|^2/L$.
This is the characteristic signature of quantum spin ice~\cite{shannon_quantum_2012,sikora_quantum_2009,sikora_extended_2011}.
\autoref{fig:ring_energy} shows the quadratic cost as a function of $\phi_x$ and the reciprocal system-size dependence.
We have set the flux through the other surfaces to zero, $\phi_y=\phi_z=0$.

\section{Package structure and API}
\label{sec:package}
The package is organized around five main components, which we describe in turn:
\begin{list}{}{%
	\setlength{\leftmargin}{0pt}%
	\setlength{\rightmargin}{0pt}%
	\setlength{\labelwidth}{0pt}%
	\setlength{\labelsep}{0pt}%
	\setlength{\itemindent}{0pt}%
}
	\item[] \textbullet\ \autoref{sec:pkg:hamiltonian}: Hamiltonian;
	\item[] \textbullet\ \autoref{sec:pkg:walker}: Initial walker ensemble;
	\item[] \textbullet\ \autoref{sec:pkg:guide}: Guiding wavefunction (optional);
	\item[] \textbullet\ \autoref{sec:pkg:obs}: Observables (optional);
	\item[] \textbullet\ \autoref{sec:pkg:driver}: Simulation driver.
\end{list}
\PyGFMC{} can be installed with \textbf{\pyinline{pip install pygfmc}} (\textbf{\pyinline{pip install "pygfmc[netket]"}} if you want to use NetKet~\cite{SciPostPhysCodeb.7,SciPostPhysCodeb.7-r3.4}), or from the source code at \repo.

\subsection{Hamiltonian}\label{sec:pkg:hamiltonian}
The Hamiltonian determines the walker propagation.
We represent any operator as a sum
\begin{align}
	H = \sum_{i=0}^{\texttt{M\_total} - 1} h_i\,,\label{eq:H}
\end{align}
where each term $h_i$ produces \emph{at most one} new state when acting on a basis state $\ket{x}$.
Operators like $S^x$ therefore must be split into separate operators, such as $S^+$ and $S^-$ for spin $S > 1/2$.

As a minimal example, consider a nearest-neighbor Hamiltonian for spin-$1/2$ on a one-dimensional chain with periodic boundary conditions, non-uniform coupling amplitudes $c_j^{z/\pm}$, and local $z$-fields $m_j$:
\begin{align}
	H \;=\; \underbrace{-\sum_{j=0}^{L-1} m_j\, S_j^z + \sum_{j=0}^{L-1}c_j^z
	S^z_{j}\, S^z_{(j+1) \bmod L}}_{\equiv\, D}
	\;-\; \sum_{j=0}^{L-1} c_j^\pm \left(
		S^+_{j}\, S^-_{(j+1) \bmod L} + \text{h.c.}\right).
	\label{eq:H_example}
\end{align}
We absorb \emph{all} diagonal operators into one single operator $D$.

A Hamiltonian class must include the following attributes and methods:

\begin{itemize}
	\item \pyinline{M_total}: total number of terms in the sum of \autoref{eq:H}.
	For the example in \autoref{eq:H_example}, $\Mtotal = L + 1$ (one diagonal + $L$ hopping bonds).

	\item \pyinline{get_rates(config, return_configs=False)}: given a configuration \pyinline{config}, returns the pair \pyinline{(rates, new_configs)}.
	\pyinline{rates} is a one-dimensional array of length \pyinline{M_total} containing the weight produced by each $h_i$.
		When \pyinline{return_configs=True}, \pyinline{new_configs} contains the configurations in an array of shape \pyinline{(M_total, N)} obtained by applying each $h_i$ to \pyinline{config}.
		It is \pyinline{None} otherwise.

	Taking the example Hamiltonian in \autoref{eq:H_example} for $L = 4$ and $\pyinline{config} = \ket{1100}$, the weight array is
	\begin{align}
		\pyinline{rates}=\begin{pmatrix}
			d &
			0 & -c_1^\pm & 0 & -c_3^\pm
		\end{pmatrix}\,,\label{eq:weights}
	\end{align}
	with $d=\tfrac{1}{2}(m_0 + m_1 - m_2 - m_3) + \tfrac{1}{4}(c_0^z-c_1^z+c_2^z-c_3^z)$ as the diagonal element.

	The first entry is \emph{always} reserved for the diagonal operator.
	It is zero if no diagonal terms are present.
	The two interior zeros come from hopping operators acting on aligned neighbors, as those bonds cannot flip and, therefore, do not generate a transition.  
	The	corresponding new configurations form a two-dimensional array, where row $i$ corresponds to applying $h_i$ to \pyinline{config},
	\begin{align}
		\pyinline{new_configs}=\begin{pmatrix}
			1 & 1 & 0 & 0 \\
			1 & 1 & 0 & 0 \\
			1 & 0 & 1 & 0 \\
			1 & 1 & 0 & 0 \\
			0 & 1 & 0 & 1
		\end{pmatrix}\,.
		\label{eq:configurations}
	\end{align}
	Rows for diagonal or non-flippable operators are not considered.
	They simply repeat the input \pyinline{config} in this case.

	\item \pyinline{generate_state(config, idx)}: returns the configuration corresponding to row \pyinline{idx} of \autoref{eq:configurations}.
		The propagator calls it whenever \pyinline{get_rates} was not asked to materialize the successor configurations.
		We recommend \emph{not} computing all \pyinline{M_total} possible configurations, since only the one stochastically chosen is required.
		This saves memory and runtime.

\item \pyinline{diagonal_weight(config)} (optional): returns the diagonal contribution $d$, which is used to compute the diagonal expectation values.
		It can also return multiple values via an array, which is beneficial if a set of diagonal operators shall be evaluated at once (see \autoref{sec:pkg:obs}).
\end{itemize}

\paragraph{Implemented Hamiltonian classes.} The following classes are implemented:
\begin{itemize}
	\item \pyinline{Spin.create(edges, S, J_z, J_pm, J_pmpm, h_z, h_pm, N)}: spin-$S$ Hamiltonian with one- and two-body operators:
	\begin{align}
		H={}&\sum_{\langle i,j\rangle}\Big[J_z S_i^zS_j^z+J_{\pm}\big(S_i^+S_j^-+{\rm h.c.}\big)+J_{\pm\pm}\big(S_i^+S_j^++{\rm h.c.}\big)\Big]\notag\\
		&-\sum_i\Big[h_zS_i^z+h_{\pm}\big(S_i^++S_i^-\big)\Big].
		\label{eq:H_spin_generic}
	\end{align}

	\item \pyinline{SpinHalf.create(edges, J_x, J_y, J_z, h_z, h_x, N)}: optimized XYZ Hamiltonian for $S = 1/2$ (used in \autoref{sec:ex:tfim} and \autoref{sec:ex:pyro_nn}):
	\begin{align}
		H=\sum_{\langle i,j\rangle}\Big[J_xS_i^xS_j^x+J_yS_i^yS_j^y+J_zS_i^zS_j^z\Big]-\sum_i\Big[h_zS_i^z+h_xS_i^x\Big].
		\label{eq:H_spinhalf_generic}
	\end{align}

	\item \pyinline{ToricCode.create(Lx, Ly, A_s, B_p, h_x, h_z)}: toric code Hamiltonian with local $x$ and $z$ fields:
	\begin{align}
		H=-A_s\sum_v\prod_{i\in\star(v)}\sigma_i^x-B_p\sum_p\prod_{i\in\Box(p)}\sigma_i^z-h_z\sum_i\sigma_i^z-h_x\sum_i\sigma_i^x.
		\label{eq:H_toric_generic}
	\end{align}

	\item \pyinline{BoseHubbard.create(edges, t, eta, U, mu, h_pm, N, n_max)}: generic Bose--Hubbard Hamiltonian:
	\begin{align}
		H={}&-t\sum_{\langle i,j\rangle}\big(b_i^\dagger b_j+{\rm h.c.}\big)
		 -\eta\sum_{\langle i,j\rangle}\big(b_i^\dagger b_j^\dagger+{\rm h.c.}\big)\notag\\
		&+\sum_i\Big[\tfrac{U}{2}n_i(n_i-1)-\mu n_i-h_{\pm}\big(b_i^\dagger+b_i\big)\Big].
		\label{eq:H_bosehubbard}
	\end{align}

\item \pyinline{PlaquetteModel.create(plaquettes, g, mu)}: implements ring-exchange flips of amplitude $-g$ on a collection of even-length plaquettes, with a Rokhsar--Kivelson potential~\cite{rokhsar_superconductivity_1988} (used in \autoref{sec:ex:ring}).
		The \pyinline{plaquettes} are simply given by a two-dimensional array containing the sites of each plaquette.
		We define $\ket{\cw_p}$ and $\ket{\ccw_p}$ for the two \emph{flippable} configurations on a plaquette $p$ where spins alternate $+,-,+,-,\dots$ clockwise or counterclockwise:
	\begin{align}
		H=-g\sum_{p}\Big(\ket{\ccw_p}\!\bra{\cw_p}+\ket{\cw_p}\!\bra{\ccw_p}\Big)
		 +\mu\sum_p\Big(\ket{\cw_p}\!\bra{\cw_p}+\ket{\ccw_p}\!\bra{\ccw_p}\Big).
		\label{eq:H_plaquette_generic}
	\end{align}
	The off-diagonal term flips a plaquette between its two alternating	configurations and the potential counts the number of flippable plaquettes.
\end{itemize}

A simple code example to construct the Hamiltonian from \autoref{eq:H_example} is:
\begin{lstlisting}
import numpy as np
from pygfmc.hamiltonian import SpinHalf

L     = 8
edges = [(i, (i + 1) % L) for i in range(L)]           # 1D PBC chain

J_xy  = -np.random.uniform(0.8, 1.2, size=len(edges))  # random per-bond
m     = np.linspace(-0.3, 0.3, L)                      # local z-fields m_j

H = SpinHalf.create(edges,
                    J_x=J_xy, J_y=J_xy,    # random per-edge XY
                    J_z=1.0,               # uniform Ising coupling
                    h_z=m)                 # linear ramp
\end{lstlisting}

\paragraph{Exact diagonalization.} \PyGFMC{} provides an option to compute the ground state (and excited states) using exact diagonalization:
\begin{lstlisting}
energies, diag_obs, off_diag_obs = exact_diagonalization(
    hamiltonian,                # any class following the rate protocol
    local_hilbert_space=None,   # per-site values, defaults to hamiltonian.local_hilbert_space
    N=None,                     # number of sites, defaults to hamiltonian.N
    diag_operators=None,        # measured as sum_x |psi_x|^2 O(x)
    off_diag_operators=None,    # measured as psi^T O psi
    number_of_states=1,         # how many lowest states to solve for
)
\end{lstlisting}
It relies on the \pyinline{get_rates} function of the Hamiltonian.
The local Hilbert space for \pyinline{N} sites is given by \pyinline{local_hilbert_space}, which must match the convention of the Hamiltonian, e.g., \pyinline{(-1,1)} for spin-1/2.
Besides the ground-state energy, it returns observables for diagonal and off-diagonal operators in a dictionary (see \autoref{sec:pkg:obs}).

More details, including a custom Hamiltonian class, can be found in \autoref{app:hamiltonians} and \autoref{app:user-defined}.\\

\subsection{Walker ensemble}\label{sec:pkg:walker}
The walker ensemble is stored as a two-dimensional array of shape $\nwalker \times N$, where $N$ is the system size.
Entry $[i, j]$ holds the basis-state value, such as a $\sigma^z$ eigenvalue or an occupation number, of the $i$-th walker at site $j$.
Any encoding can be used, as long as it is consistent with the Hamiltonian and the (optional) guiding wavefunction.

Note that the initial walker ensemble must be chosen such that every walker lies within the symmetry sector of interest, for example, a sector defined by fixed total particle number or magnetization.
We use this in \autoref{sec:ex:ring} to target topological electric-flux sectors.
If the initial walker ensemble is spread across different sectors, it samples the different ground states within these symmetry sectors, and the simulation is corrupted.\\

\PyGFMC{} provides several routines for initializing the walker ensemble:
\begin{itemize}
\item \pyinline{prepare_walkers_random(N_walkers, N_sites, rng_key, states=(-1,1))}: each site is drawn independently and uniformly from \pyinline{states}, which encodes, for example, spin-$1/2$ degrees of freedom by setting it to \pyinline{(-1,1)} or \pyinline{(-0.5,0.5)}, higher-spin values to \pyinline{(-1,0,1)}, or bosonic occupation numbers to $(0,\ldots,n_{\max})$.
	\item \pyinline{prepare_walkers_sz(N_walkers, N_sites, rng_key, sz=0, states=(-1,1))}: random configurations at a fixed total charge.
		\pyinline{states} defines the local Hilbert space and \pyinline{sz} simply refers to the sum over all sites $\sum_i\sigma^z_i$.
	\item \pyinline{prepare_walkers_from_configs(configs)}: the walkers are initialized with the given configurations \pyinline{configs} of shape \pyinline{(N_walkers, N_sites)}.
\end{itemize}
The configuration's data type is determined by \pyinline{states} or \pyinline{configs}.

We note that randomly generated initial states are usually poor representations of the ground-state ensemble, so a warm-up should be executed before the main pipeline.\\

Continuing the example code for the XXZ chain of \autoref{eq:H_example}, which conserves the total magnetization, we initialize the walkers in the $S^z_{\rm tot} = 0$ sector.
\begin{lstlisting}
import time, jax
from pygfmc.walkers import prepare_walkers_sz

rng       = jax.random.PRNGKey(int(time.time()))   # time-seeded
rng, key  = jax.random.split(rng)
walkers   = prepare_walkers_sz(N_walkers=512, N_sites=H.N, rng_key=key, sz=0)
\end{lstlisting}

\subsection{Guiding wavefunction (optional)}\label{sec:pkg:guide}
{\emergencystretch=3em
The algorithm requires only the ratio $\psi_T(x')/\psi_T(x)$, so the guide class returns the log-difference $\log \psi_T(x') - \log \psi_T(x)$ for all \pyinline{M_total} successor states of the current configuration.
At least one of the following three methods must be implemented:
\begin{itemize}
	\item \pyinline{log_diff_from_operator(config, hamiltonian, cache=None)}: evaluates the log-differences directly from \pyinline{config} and the \pyinline{hamiltonian}.
		This function does not require the materialized \pyinline{M_total} configurations and computes the logarithmic difference on the fly.
		This saves memory and runtime.

	\item \pyinline{log_diff_from_configs(config, new_configs, cache=None)}: takes the pre-materialized \pyinline{new_configs} array from \pyinline{get_rates(config, return_configs=True)} (see \autoref{eq:configurations}).
		This function is used as a fallback when the previous method is not available.

	\item \pyinline{log_psi(config, cache=None)}:
		evaluates the log-amplitude $\log\psi_T$ directly for a given \pyinline{config}.
		It is used as a last
	fallback when neither of the other two methods is available.
\end{itemize}
}

The guide may optionally provide a function \pyinline{build_cache(hamiltonian)}, which returns a \pyinline{cache} object that is passed to the methods above.
For example, it can store system-specific information, which often helps to speed up the calculation.\\

\PyGFMC{} has different pre-implemented guides: 
\begin{itemize}
	\item \pyinline{JastrowSpinHalf(N, pairs, beta, h=None)} for \pyinline{SpinHalf}: the pairwise ansatz used in \autoref{eq:jastrow_TFIM}
		\begin{align}
			\psi_T(x) \;=\; \exp\!\Big(
				-\sum_{\langle i,j\rangle} \beta_{ij}\, \sigma^z_i\, \sigma^z_j
				+\sum_i h_i\, \sigma^z_i
			\Big)\,,
			\label{eq:jastrow}
		\end{align}
		where $\beta_{ij}$ ($h_i$) can vary across spin pairs (sites) or be a single global parameter.
	\item \pyinline{JastrowSpin(N, pairs, beta, h=None, d=None)} for \pyinline{Spin}: similar to the ansatz in \autoref{eq:jastrow}:
		\begin{align}
			\psi_T(x) \;=\; \exp\!\Big(
				-\sum_{\langle i,j\rangle} \beta_{ij}\, x_i\, x_j
				+\sum_i h_i\, x_i
				-\sum_i d_i\, x_i^2
			\Big)\,,
			\label{eq:spinjastrow}
		\end{align}
		where $x_i \in \{-S,\dots,S\}$ is the eigenvalue of $S^z_i$ and $d_i$ is a single-ion term.
	\item \pyinline{JastrowBoseHubbard(N, pairs, beta, h=None, d=None)} for \pyinline{BoseHubbard}: similar to the ansatz in \autoref{eq:spinjastrow}:
		\begin{align}
			\psi_T(x) \;=\; \exp\!\Big(
				-\sum_{\langle i,j\rangle} \beta_{ij}\, x_i\, x_j
				+\sum_i h_i\, x_i
				-\sum_i d_i\, x_i^2
			\Big)\,,
			\label{eq:bosonjastrow}
		\end{align}
		where $x_i$ is the occupation of site $i$ and $d_i$ is the Gutzwiller factor.

	\item \pyinline{ClusterCPS(groups, amplitudes=None, init='zero', seed=None)} for \pyinline{SpinHalf}: a \emph{correlator product state}~\cite{changlani_approximating_2009,mezzacapo_ground-state_2009}, which assigns one variational parameter to every configuration of a local cluster, $\Phi_c(x_c) \mapsto \mathbb{R}$, using a lookup table.
		Here $x_c$ is the local configuration on a cluster $c$.
		Symmetries can be embedded, making local clusters of up to $18$ sites possible.
		This ansatz has been used in \autoref{sec:ex:pyro_nn}.
	\item \pyinline{ClusterMLP(groups, params=None, seed=None)} for \pyinline{SpinHalf}: implements the cluster functions $\Phi_c$~\cite{changlani_approximating_2009,mezzacapo_ground-state_2009} as multilayer perceptrons (MLPs) with a configurable number of hidden layers.

\end{itemize}

\PyGFMC{} further provides a generic function to optimize a guide with \texttt{NetKet}~\cite{SciPostPhysCodeb.7,SciPostPhysCodeb.7-r3.4}:
\begin{lstlisting}
guide = train_with_netket(
    hamiltonian,                # only get_rates is required
    guide,                      # only log_psi is required
    local_hilbert_space=None,   # defaults to hamiltonian.local_hilbert_space
    N=None,                     # defaults to hamiltonian.N
    ...,                        # sampler and further VMC settings
)
\end{lstlisting}
A new guide of the same class is returned and the original is left untouched.\\

Continuing the running example, a one-parameter Jastrow ansatz $\beta_{ij} = \beta$ defined on the Hamiltonian edges can be created via:
\begin{lstlisting}
from pygfmc.wavefunction import JastrowSpinHalf

guide = JastrowSpinHalf(H.N, H.edges, beta=0.1)
\end{lstlisting}
Further implementation details are given in the appendix: implemented guides (\autoref{app:guides}), user-defined guides (\autoref{app:user-defined-guides}), and the \texttt{NetKet}~\cite{SciPostPhysCodeb.7,SciPostPhysCodeb.7-r3.4} integration (\autoref{app:netket}).

\subsection{Observables (optional)}\label{sec:pkg:obs}
An observable can be created with the same interface as the Hamiltonian.
While an off-diagonal observable needs \pyinline{M_total} and \pyinline{get_rates(config)}, as a Hamiltonian does, a diagonal observable needs only \pyinline{diagonal_weight(config)}, which is cheaper to evaluate.
As discussed in \autoref{sec:gfmc:obs}, the matrix elements of the off-diagonal operator must be nonnegative.
Operators with mixed signs must be split into positive and negative parts and measured separately (see \autoref{eq:observable_splitting}).
The observables are passed to the main pipeline as dictionaries keyed by name, via the \pyinline{diag_operators} and \pyinline{off_diag_operators} arguments (see \autoref{sec:pkg:driver}), and the returned estimates are mirrored back in the same key structure.\\

Launching an auxiliary Markov chain to measure an off-diagonal observable is expensive and scales with $m_{\rm max}=\max(p)/2$.
It is therefore worth determining the autocorrelation time first and launching the chain only once the walker ensemble has decorrelated.
Within the production run, this is controlled by the \pyinline{off_diag_stride} keyword, which measures the off-diagonal observables only every \pyinline{off_diag_stride}-th branching step.
The energy and the diagonal observables are unaffected and remain resolved at every step.\\

\PyGFMC{} allows \pyinline{diagonal_weight(config)} to return multiple values, which can be beneficial in some cases.
This is the case for the diagonal $S^zS^z$ structure factor that can be computed for several momenta simultaneously via \pyinline{SzSzStructureFactor(positions, k_vectors)} for the \pyinline{SpinHalf} class.
The function takes the positions of the sites, given as an array of shape $(N,\mathrm{dim})$, and a selection of $k$ vectors of the same dimension:
\begin{align}
	S^{zz}(\mathbf{k})= \sum_{i<j} 2\cos\!\big(\mathbf{k}\cdot\left(\mathbf{r}_{i}-\mathbf{r}_{j}\right)\big)\,S^z_iS^z_j\,.\label{eq:szsz_structure_factor}
\end{align}
The $\mathbf{k}$ vectors must be compatible with the periodic boundary conditions, $\mathbf{k}\cdot\mathbf{a}_i\in 2\pi\mathbb{Z}$ for the lattice vectors $\mathbf{a}_i$ spanning the system.

\pyinline{StructureFactor(positions, k, channel='SpSm')} returns the corresponding off-diagonal operator for the transverse structure factor of the \pyinline{SpinHalf} class,
\begin{align}
	{O} =& \sum_{i<j}\Big[1+\cos\!\big(\mathbf{k}\cdot\left(\mathbf{r}_{i}-\mathbf{r}_{j}\right)\big)\Big]
	\Big(A_iB_j+\mathrm{h.c.}\Big)
	={AB}(0)+{AB}(\mathbf{k}),\\
	\mathrm{with }\qquad{AB}(\mathbf{k})=&\sum_{i<j}\cos\!\big(\mathbf{k}\cdot\left(\mathbf{r}_{i}-\mathbf{r}_{j}\right)\big)\Big(A_iB_j+\mathrm{h.c.}\Big)\,.
\end{align}
There are different options for the channel: \pyinline{'SpSm'}: $A=S^+$ and $B=S^-$; \pyinline{'SpSp'}: $A=S^+$ and $B=S^+$; \pyinline{'SpSm+SpSp'}: $A=B=\sqrt{2} S^x$, which results in $A_iB_j+\mathrm{h.c}=S^+_iS^-_j+S^+_iS^+_j+\mathrm{h.c.}=4S^x_iS^x_j$.
The channel \pyinline{'SzSz'} ($A=B=S^z$) yields the diagonal operator $S^{zz}(0)+S^{zz}(\mathbf{k})$ of \autoref{eq:szsz_structure_factor}.
Notably, we compute the structure factor at a given $k$ vector plus the zero vector.
This avoids the problem that the cosine becomes negative.\\

We demonstrate how to compute the total and on-site magnetization operators $S^z_{\rm tot} = \sum_i S^z_i$ and $S^z_0$, their transverse counterparts $S^x_{\rm tot}$ and $S^x_0$, and the diagonal and transverse structure factors.
We set \pyinline{sign_check=False}, which disables the sign-problem-free check that is enforced for Hamiltonians but not needed for observables.
\begin{lstlisting}
import jax.numpy as jnp
from pygfmc.hamiltonian import SpinHalf, SzSzStructureFactor, StructureFactor

h_all = jnp.ones(L)
h_0   = jnp.zeros(L).at[0].set(1.0)

Sztot = SpinHalf.create(edges=[], N=L, h_z=-h_all)
Sz0   = SpinHalf.create(edges=[], N=L, h_z=-h_0)
Sxtot = SpinHalf.create(edges=[], N=L, h_x=-h_all, sign_check=False)
Sx0   = SpinHalf.create(edges=[], N=L, h_x=-h_0,   sign_check=False)

# structure factors: all allowed k of the L-site ring at once
positions = jnp.arange(L)[:, None]                    # (N, dim)
k_vectors = 2 * jnp.pi * jnp.arange(L)[:, None] / L   # (n_k, dim)

Szz = SzSzStructureFactor(positions, k_vectors)
Spm = StructureFactor(positions, k_vectors[L // 2], channel='SpSm')

diag_operators     = {'Sztot': Sztot, 'Sz0': Sz0, 'Szz': Szz}
off_diag_operators = {'Sxtot': Sxtot, 'Sx0': Sx0, 'Spm': Spm}
\end{lstlisting}
\autoref{app:user-defined} demonstrates how to implement custom operators that can be fed into the \PyGFMC{} pipeline, such as multi-body terms or a simplified version of the structure factor.

\subsection{Simulation driver}\label{sec:pkg:driver}
The four components above (Hamiltonian, walkers, guide, and observables) are tied together using the \pyinline{run} function, which is the core of \PyGFMC{}.
It executes an optional warm-up, the production scan of $\nsteps$ branching steps, and the post-run statistics.
When several GPUs are visible, \pyinline{run} automatically distributes the walker population across all of them.
Measuring diagonal observables is restricted to a single device.

\paragraph{Key input keywords.}
A complete list of the input arguments of \pyinline{run} is provided in \autoref{app:run_input}.
The most important keywords are:
\begin{apilistbox}
\begin{itemize}
	\item \pyinlinebold{walkers}: the initial walker ensemble (\autoref{sec:pkg:walker});
	\item \pyinlinebold{n_steps}: number of production branching steps;
	\item \pyinlinebold{hamiltonian}: the Hamiltonian;
	\item \pyinlinebold{step_size}: an integer number of steps for the discrete methods (\pyinline{"single_step"}, \pyinline{"multi_step"}), or a positive real imaginary-time length~$\beta$ for \pyinline{"cont_time"} (see \autoref{sec:gfmc:cont_time});
	\item \pyinlinebold{guide}: guiding wavefunction (use \pyinline{None} if no guide should be used);
	\item \pyinlinebold{p_array}: list containing the number of branching steps $p$ defining the projection depth (see \autoref{sec:gfmc:params});
	\item \pyinlinebold{method}: propagation scheme (see \autoref{sec:gfmc:cont_time}):
			\begin{itemize}
	      	\item \pyinline{"single_step"} for step-by-step propagation,
	      	\item \pyinline{"multi_step"} for batched discrete substeps, or
	      	\item \pyinline{"cont_time"} for the continuous-time scheme.
			\end{itemize}
	       (Default: \pyinline{"cont_time"});
	   \item \pyinlinebold{rng_key}: \JAX{} PRNG key (Default: \pyinline{jax.random.PRNGKey(secrets.randbits(32))});
	   \item \pyinlinebold{lam}: diagonal shift~$\lambda$ (only required for discrete propagation methods);
	   \item \pyinlinebold{diag_operators}: diagonal observables as a keyed \texttt{dictionary} as described in \autoref{sec:pkg:obs} (Default: \pyinline{None});
	   \item \pyinlinebold{off_diag_operators}: off-diagonal observables as a keyed \texttt{dictionary} as described in \autoref{sec:pkg:obs} (Default: \pyinline{None}).
\end{itemize}
\end{apilistbox}

\Needspace{7\baselineskip}
\paragraph{Return values.} The function returns the tuple \pyinline{(walkers, energy, obs, stats, raw)} (raw is optional), where
\begin{apilistbox}
\begin{itemize}
	\item \pyinlinebold{walkers}: the updated walker batch;
	\item \pyinlinebold{energy}: a \pyinline{dict} of length \pyinline{len(p_array)};
	\item \pyinlinebold{obs}: a nested \pyinline{dict} accessed as \pyinline{obs[name][p]}, where the outer key is the name of the operator and the inner key is the number of projection steps $p$;
	\item \pyinlinebold{stats}: a dictionary with the attributes of the \pyinline{statistics} function described in \autoref{sec:gfmc:simulation}, together with \pyinline{'avg_log_weight'}, \pyinline{'it_per_sec'}, \pyinline{'runtime'} and \pyinline{'total_wall_time'}, the autocorrelation times under the \pyinline{'autocorrelation/'} prefix, and the run's own inputs under \pyinline{'parameters/'};
	\item \pyinlinebold{raw}: a dictionary with the raw, step-resolved GFMC values: \pyinline{log_weights} and \pyinline{energy} are 1D arrays of length $\nsteps$.
		Observables are accessed via nested keys with \pyinline{['diag']['<name>']} and \pyinline{['off_diag']['<name>']} that contain arrays of shape $\nsteps \times n_p$, where $n_p$ is the length of \pyinline{p_array}.
		Diagonal operators may hold an array of shape $\nsteps \times n_k \times n_p$, where $n_k$ is the number of return values of \pyinline{diagonal_weight}.
		$\nsteps$ is reduced if \pyinline{off_diag_stride} is greater than 1 for the off-diagonal observable.
		They contain the raw estimators of the observables.
		The dictionary also includes \pyinline{off_diag_norm}, which has the same dimensions and contains the logarithmic weight obtained by the auxiliary chain evolved over $m=p/2$ steps.
\end{itemize}
\end{apilistbox}

\paragraph{Storing and resuming a simulation.} Objects returned from the production run can simply be stored using the function:
\begin{lstlisting}
write_to_HDF5(
    filename,              # created if absent, appended to otherwise
    stats,                 # statistics and input parameters, required
    walkers=None,          # final walker ensemble
    energy=None,           # energy per p
    obs=None,              # observables per p
    raw=None,              # per-step energy, log weights and observables
    walker_filename=None,  # separate file for the walker ensemble
)
\end{lstlisting}

It is designed both to create a file initially and to append data from consecutive simulation runs.
Data are categorized into four groups:
\begin{itemize}
	\item[(i)] The initial creation stores all fixed input parameters, including data types and hashes for the Hamiltonian, guide, and observables.
	These are stored in the group \pyinline{'/parameters'}.
	When a simulation is resumed (see below), for example, when more steps are required for better statistics, all relevant parameters are checked to ensure consistency between simulations.
	\item[(ii)] Each simulation also stores data in a designated group, \pyinline{'/run_<n>'} for the $n$-th simulation.
	This group contains the following (when provided): the energy per $p$ (\pyinline{energy}), the observables per $p$ (\pyinline{obs}), and statistics (\pyinline{stats}).
	The statistics include the branching diagnostics of \autoref{sec:gfmc:simulation}, the autocorrelation times, and other information such as the iterations per second or the runtime.
	We further store a hash of the final walker ensemble and the seed to ensure a continuous, reproducible simulation.
	\item[(iii)] The final walker ensemble (needed to relaunch the simulation) is stored in \pyinline{'/current_walkers'} if provided.
	Storing the walker ensemble might require a large amount of disk space.
	Hence, we provide the option to store it instead in \pyinline{walker_filename} when it is provided.
	It is overwritten with each new call.
	\item[(iv)] The raw data (per-step energy estimator, logarithmic weights, or observables) are appended in the group \pyinline{'/raw'}.
	These arrays continuously grow in size whenever another production run is launched.
\end{itemize}

The simulation can be relaunched from the HDF5 file using
\begin{lstlisting}
walkers, energy, obs, stats, raw = run_from_HDF5(
    filename,                   # file written by write_to_HDF5
    hamiltonian,                # checked against the stored hash
    guide=None,                 # checked against the stored hash
    diag_operators=None,        # checked against the stored hashes
    off_diag_operators=None,    # checked against the stored hashes
    n_steps=None,               # None reuses the last number of steps
    walker_filename=None,       # read the ensemble from a separate file
)
\end{lstlisting}
This is essentially a wrapper for the \pyinline{run} function and returns the same objects.
Before it launches the calculation, it verifies that the input classes (Hamiltonian, observables, and guide) have matching hashes.
Since the simulation is relaunched from an equilibrated ensemble, the warm-up is skipped.
When \pyinline{n_steps} is not given, it reuses the number of steps used in the previous calculation.

The file grows continuously and holds the entire main chain. Continuing with the stored RNG key reproduces the trajectory of one contiguous run.
Therefore, it can be used to resume a long simulation after a timeout or crash when intermediate steps are stored.\\

\PyGFMC{} can compute observables and energy from the \pyinline{raw} dictionary, which is either returned from the \pyinline{run} function or loaded from disk via \pyinline{raw = load_raw_from_HDF5(filename)}:
\begin{lstlisting}
energy, obs = evaluate_observables(
    raw,             # raw data
    p_array=None,    # None reuses the stored branching steps
    n_blocks=None,   # number of jackknife blocks; None returns no errors
)
\end{lstlisting}
The energy can be assessed for arbitrary branching steps in post-processing by passing an array of $p$ values to the function.
This is not the case for observables, as they have already been computed for a given number of steps $p$.

The function can also return estimates of the error using a blocked jackknife as discussed in \autoref{sec:errors}.
If \pyinline{n_blocks} is given, the function returns \pyinline{(energy, obs, energy_err, obs_err)} instead.
However, we recommend running independent simulations and comparing their means rather than relying solely on the jackknife analysis.
Note that both approaches estimate only the statistical error and do not capture the systematic bias from a finite projection depth.\\

We complete our example with:
\begin{lstlisting}
from pygfmc import run, write_to_HDF5, run_from_HDF5

walkers, energy, obs, stats, raw = run(
    walkers, n_steps=2000, hamiltonian=H, step_size=0.5,
    guide=guide, p_array=[4, 8, 16, 32, 48], warmup_steps=500,
    method='cont_time', rng_key=rng,
    diag_operators=diag_operators,
    off_diag_operators=off_diag_operators,
)
write_to_HDF5('run.h5', stats, walkers, energy, obs, raw)

# continue the same chain for another 2000 steps
walkers, energy, obs, stats, raw = run_from_HDF5(
    'run.h5', H, guide, diag_operators, off_diag_operators,
)
write_to_HDF5('run.h5', stats, walkers, energy, obs, raw)
\end{lstlisting}
Parts of the code snippets above are collected in \texttt{notebooks/xxz\_chain.ipynb}.

\section{Discussion}\label{sec:discussion}
\PyGFMC{} provides a GPU-accelerated implementation of Green's function Monte Carlo for sign-problem-free quantum lattice models.
By exploiting the substantial computational capabilities of modern GPUs, it can study problems that were previously hardly accessible.
The resulting speedup of nearly three orders of magnitude enables more expressive guiding wavefunctions and walker populations of up to $10^7$ (see \autoref{fig:throughput}).
We have demonstrated that \PyGFMC{} reliably resolves the ground state of frustrated magnets for which other quantum Monte Carlo methods, such as the stochastic series expansion, struggle to reach low temperatures~\cite{kato_numerical_2015,huang_extended_2020}.
This makes it a competitive alternative for ground-state properties.

The library provides built-in Hamiltonians and guiding wavefunctions, diagonal and off-diagonal estimators, HDF5-based storage and continuation, and support for custom classes without modifying the core engine.
Furthermore, it integrates with \texttt{NetKet}~\cite{SciPostPhysCodeb.7,SciPostPhysCodeb.7-r3.4}, providing access to a broad range of variational wavefunctions and robust training procedures.
The repository includes various notebooks with examples, and detailed documentation is available at \repo{}.

The code was developed from our earlier \texttt{C++} implementation with the assistance of coding tools based on large language models and has been validated through extensive test suites (see \autoref{sec:pkg:tests}). 
The manuscript, documentation, and code comments are designed to make the library accessible to both human users and coding agents.
We encourage agent-assisted development on top of \PyGFMC{}.

We see two directions for future work.
First, while GFMC with its fixed walker ensemble is especially well suited to GPUs, other variants of quantum Monte Carlo (see Refs.~\cite{kim_qmcpack_2018,bernaschi_the_2024}) would likely benefit from the same computational power.
Thus, it is worth exploring GPU implementations of other quantum Monte Carlo methods.
Second, we find that a carefully chosen ansatz wavefunction is the key to a well-converged simulation.
Implementing and benchmarking further variational wavefunctions, such as physically motivated neural quantum states~\cite{kufel_approximately_2025}, therefore offers a promising route to increasingly challenging problems such as spin glasses.

\section*{Acknowledgments}
Many thanks to David Luitz, Nils Niggemann, and Johannes Reuther for helpful discussions and feedback.

\paragraph{Funding information}
This work was funded by the DFG under Project No.~575641691.

\bibliography{PyGFMC}

\appendix
\section{Arguments for the production run}\label{app:run_input}
The production \pyinline{run} takes the following arguments:
\begin{itemize}
	\item \pyinline{walkers}: the initial walker ensemble (\autoref{sec:pkg:walker});
	\item \pyinline{n_steps}: number of production branching steps;
	\item \pyinline{hamiltonian}: the Hamiltonian;
	\item \pyinline{step_size}: an integer number of steps for the discrete methods (\pyinline{"single_step"}, \pyinline{"multi_step"}), or a positive real imaginary-time length~$\beta$ for \pyinline{"cont_time"} (see \autoref{sec:gfmc:cont_time});
	\item \pyinline{guide}: guiding wavefunction (use \pyinline{None} if no guide should be used);
	\item \pyinline{p_array}: list containing the number of branching steps $p$ defining the projection depth (see \autoref{sec:gfmc:params});
	\item \pyinline{method}: propagation scheme (see \autoref{sec:gfmc:cont_time}):
			\begin{itemize}
	      	\item \pyinline{"single_step"} for step-by-step propagation,
	      	\item \pyinline{"multi_step"} for batched discrete substeps, or
	      	\item \pyinline{"cont_time"} for the continuous-time scheme.
			\end{itemize}
	       (Default: \pyinline{"cont_time"});
	\item \pyinline{rng_key}: \JAX{} PRNG key (Default: \pyinline{jax.random.PRNGKey(secrets.randbits(32))});
	\item \pyinline{lam}: diagonal shift~$\lambda$ (only required for discrete propagation methods);
   \item \pyinline{n_devices}: number of devices used by \pyinline{run} (Default: \pyinline{None}, meaning \pyinline{jax.device_count()}).
   Set $1$ to force single-device execution.
   Runs containing diagonal observables automatically fall back to a single device;
   \item \pyinline{diag_operators}: diagonal observables as a keyed \texttt{dictionary} as described in \autoref{sec:pkg:obs} (Default: \pyinline{None});
   \item \pyinline{off_diag_operators}: off-diagonal observables as a keyed \texttt{dictionary} as described in \autoref{sec:pkg:obs} (Default: \pyinline{None});
	\item \pyinline{off_diag_stride}: measure the off-diagonal operators only every \pyinline{off_diag_stride}-th production step (see \autoref{sec:pkg:obs}) instead of at every step (Default: $1$);
	\item \pyinline{off_diag_batch_size}: number of walkers to which an off-diagonal operator is applied at once.
		This reduces the memory peak when an off-diagonal operator is applied to an ensemble by a factor of $\nwalker/\pyinline{off_diag_batch_size}$ (Default: \pyinline{None});
	\item \pyinline{warmup_steps}: total number of warm-up steps, split across four stages. If \pyinline{avg_log_weight} (see below) is not given, the warm-up determines \pyinline{avg_log_weight} (Default: \pyinline{None});
	\item \pyinline{avg_log_weight}: normalization of the weights at each branching step (see \autoref{eq:avg_weight}).
		When not given and \pyinline{warmup_steps} is not set, $1000$ warm-up steps are executed to determine it (Default: \pyinline{None});
	\item \pyinline{verbosity}: integer $0$--$10$ controlling the amount of printed output (Default: $7$);
	\item \pyinline{number_of_outputs}: number of progress lines printed during production (Default: $100$);
	\item \pyinline{return_raw}: return \pyinline{(walkers, energy, obs, stats, raw)} when \pyinline{True} (Default), or \pyinline{(walkers, energy, obs, stats)} when \pyinline{False}.
\end{itemize}
The first six arguments are positional; all others are keyword arguments.
\section{Validation}\label{sec:pkg:tests}
We have implemented several tests to verify the code within two test suites.
The first suite, \texttt{core}, checks individual parts of the code.
The second suite, \texttt{benchmark}, performs end-to-end benchmarks of various models against \texttt{QuSpin}~\cite{weinberg_quspin_2017,weinberg_quspin_2019}, \texttt{DanceQ}~\cite{schaefer_danceQ_2025,DanceQ_code}, and \texttt{TeNPy}~\cite{tenpy2024}.
The suites take about $30$ and $140$ minutes to run on a single CPU, respectively.

\subsection{Core tests}
The core suite contains over $500$ tests.
We summarize the most relevant ones below.

\paragraph{Diagonal and off-diagonal estimators.}
\begin{itemize}
	\item We test the auxiliary chain used to measure the off-diagonal operator by measuring the identity operator. 
		When the auxiliary chain is propagated with the same seed as the main Markov chain, the weights of the auxiliary chain equal those of the main Markov chain.

	\item Diagonal observables are measured both directly and with the (off-diagonal) forward-walking technique.
		The two estimates must agree to floating-point precision.

	\item We compare runs with different \pyinline{off_diag_stride} values against a run without stride.
\end{itemize}

\paragraph{Guiding wavefunctions.}
\begin{itemize}
	\item A guide may implement any of the three functions:
		\pyinline{log_psi}, \pyinline{log_diff_from_configs}, and \pyinline{log_diff_from_operator}.
		We check that all implemented methods are mutually consistent.

	\item We test consistent writing to and loading from disk using HDF5.

	\item We have implemented the option to use the exact wavefunction as a guide.
		We check that the local energy equals the ground-state energy on every configuration. 
		This tests the correct implementation of the transition rates.
		We further compute off-diagonal estimators with the exact guide.
	
	\item We test the \texttt{NetKet} training of each guide.
\end{itemize}

\paragraph{Other tests.}
\begin{itemize}
	\item For every Hamiltonian class, we probe the functions \pyinline{get_rates} and \pyinline{generate_state} against an explicit matrix construction for every configuration of a small system.
		
	\item We further check that the \pyinline{diagonal_weight} function matches the first entry of \pyinline{get_rates}.

	\item We ensure that single-device and multi-device runs reproduce the same data within statistical noise.

	\item We probe writing and loading the walker population to and from disk in HDF5.

	\item We verify the generation of different walker ensembles.

	\item We check bitwise identity by splitting a simulation into different segments using the \pyinline{write_to_HDF5} and \pyinline{run_from_HDF5} functions.
	\item A golden-output regression test pins short runs over every combination of propagator, Hamiltonian class, and compatible guide.
		We measure the observables, configurations, weights, local energies, and statistics.
\end{itemize}

\subsection{Benchmark against other methods}
We benchmark \PyGFMC{} against \texttt{QuSpin}~\cite{weinberg_quspin_2017,weinberg_quspin_2019}, \texttt{DanceQ}~\cite{schaefer_danceQ_2025,DanceQ_code}, and \texttt{TeNPy}~\cite{tenpy2024}.
\autoref{tab:validation-models} lists all tested combinations of Hamiltonian class and couplings.
For each combination, we compare the ground-state energy and available diagonal reference observables.
We further compare off-diagonal observables for the combinations marked in the last column of \autoref{tab:validation-models}.

\begin{figure}[t]
	\centering
	\includegraphics[width=\textwidth]{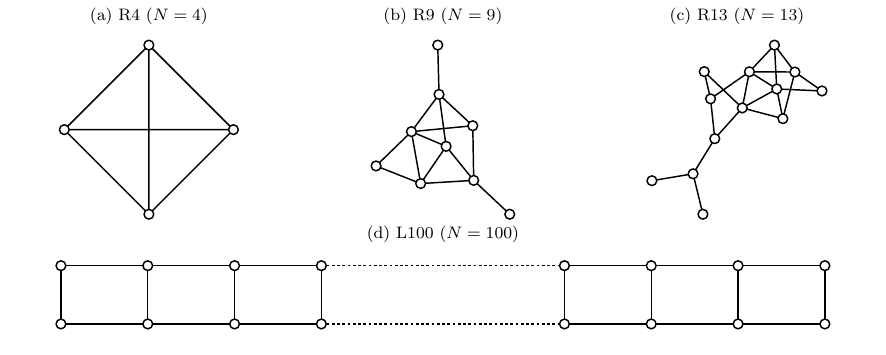}
	\caption{Geometries used in \autoref{tab:validation-models}.
		(a)--(c): The randomly generated graphs R4, R9, and R13 with $N=4$, $9$, and $13$ sites.
		(d): The two-leg ladder L100 with open boundary conditions.
	}
	\label{fig:test-lattices}
\end{figure}

The geometries we used are labeled by type and number of sites $N$.
They are partially shown in \autoref{fig:test-lattices}.
R4, R9, and R13 are randomly generated graphs with $4$, $9$, and $13$ sites, shown in \figref{fig:test-lattices}{a--c}.
T8, T12, and T16 are toric-code lattices~\cite{kitaev_fault_2003} on periodic $2\times2$, $2\times3$, and $2\times4$ square lattices, where the spins sit on the $N=2L_xL_y$ edges.
To benchmark larger systems with DMRG, we use a two-leg ladder with $L=50$ rungs and open boundary conditions (L100), see \figref{fig:test-lattices}{d}.
The ring-flip model is benchmarked on a pyrochlore lattice containing $N=128$ sites constructed from $2\times2\times2$ cubic unit cells and is labeled P128.
We compute the ground state in the zero-flux ice sector, whose Hilbert-space dimension is $8.1\times10^{10}$, using constrained exact diagonalization.
The data are obtained by a massively parallelized Lanczos calculation based on \texttt{DanceQ}~\cite{schaefer_danceQ_2025,DanceQ_code}.
All models are run with the \pyinline{"multi_step"} method.
The \pyinline{SpinHalf} sets are additionally run with \pyinline{"single_step"} and \pyinline{"cont_time"} for every geometry.

\begin{table}[htbp]
\centering
\tablefontsize
\begin{tabular}{@{}l p{6.0cm} l c@{}}
\toprule
\textbf{Hamiltonian} & \textbf{Parameters} & \textbf{Geometry} & \textbf{Off-diag.} \\
\midrule
\texttt{Spin} & $S=1$, $J_z=1$, $J_{\pm}=-1$ & R4, R9, L100 & $\checkmark$ \\
\texttt{Spin} & $S=1$, $J_z=1$, $J_{\pm}=-1$, $J_{\pm\pm}=-0.5$, $h_z=4$, $h_{\pm}=0.5$ & R4, R9, L100 & --- \\
\texttt{Spin} & $S=1$, $J_z=1$, $J_{\pm}=-1$, $h_{\pm}=2$ & R4, R9, L100 & --- \\
\texttt{SpinHalf} & $J_x=-1$, $J_y=-1$, $J_z=1$ & R4, R9, R13, L100 & --- \\
\texttt{SpinHalf} & $J_x=0$, $J_y=0$, $J_z=-4$, $h_x=4$ & R4, R9, R13, L100 & $\checkmark$ \\
\texttt{SpinHalf} & $J_x=0$, $J_y=0$, $J_z=-4$, $h_x=8$ & R4, R9, R13, L100 & --- \\
\texttt{BoseHubbard} & $t=0.5$, $\eta=0.5$, $U=4$, $\mu=2$, $n_{\max}=2$ & R4, R9, L100 & --- \\
\texttt{BoseHubbard} & $t=1$, $\eta=0$, $U=2$, $\mu=1$, $n_{\max}=2$ & R4, R9, L100 & $\checkmark$ \\
\texttt{ToricCode} & $A_s=1$, $B_p=1$, $h_z=0$, $h_x=0$ & T8, T12, T16 & $\checkmark$ \\
\texttt{ToricCode} & $A_s=1$, $B_p=1$, $h_z=0.1$, $h_x=0.1$ & T8, T12, T16 & --- \\
\texttt{ToricCode} & $A_s=1$, $B_p=1$, $h_z=0.5$, $h_x=0.5$ & T8, T12, T16 & --- \\
\texttt{ToricCode} & $A_s=1$, $B_p=1$, $h_z=2$, $h_x=2$ & T8, T12, T16 & --- \\
\texttt{PlaquetteModel} & $g=1$, $\mu=0$ & P128 & --- \\
\texttt{PlaquetteModel} & $g=1$, $\mu=1$ & P128 & $\checkmark$
 \\
\bottomrule
\end{tabular}
\caption{
	Models benchmarked against \texttt{QuSpin}~\cite{weinberg_quspin_2017,weinberg_quspin_2019}, \texttt{DanceQ}~\cite{schaefer_danceQ_2025,DanceQ_code}, and \texttt{TeNPy}~\cite{tenpy2024}.
	The first column lists the Hamiltonian class, the second lists its parameters, the third lists the underlying lattice, and the fourth marks whether off-diagonal observables were measured.
	The lattices are labeled by type and number of sites $N$: \textbf{R$N$} denotes randomly generated graphs (R4, R9, and R13; \figref{fig:test-lattices}{a--c}); \textbf{L$N$} denotes the two-leg ladder with open legs (L100; \figref{fig:test-lattices}{d}); \textbf{T$N$} denotes the toric code model on a periodic $L_x\times L_y$ torus with $N=2L_xL_y$; and \textbf{P128} denotes the pyrochlore lattice with $128$ sites.
	The R and T geometries are checked against \texttt{QuSpin} (exact diagonalization), and L100 is checked against \texttt{TeNPy} (DMRG).
	P128 is checked against a Lanczos calculation performed inside the ice manifold ($\mu=0$) and against exact analytical results at the Rokhsar--Kivelson point ($\mu=1$).
	Each row is compared with its reference for the ground-state energy and a set of diagonal observables.
	}
\label{tab:validation-models}
\end{table}

\section{Convergence in the pyrochlore XXZ model}\label{app:pyro_tau}

\begin{figure}[t]
	\centering
	\includegraphics[width=0.7\textwidth]{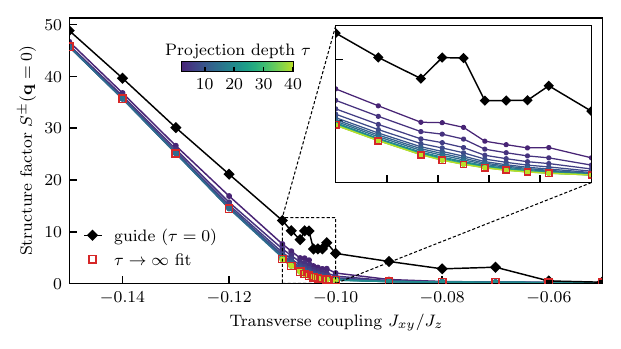}
	\caption{Convergence of $S^{\pm}(\bm{q}=0)$ from \autoref{eq:Spm_q0} with the projection depth $\tau$ for the pyrochlore XXZ model from \autoref{sec:ex:pyro_nn} at $N=1024$.
		The color bar refers to different $\tau$ values.
		Red squares show the extrapolation $\tau\to\infty$, obtained by fitting $c\,e^{-a\tau}+b$ to the data for $\tau\geq 10$.
		Black diamonds show the expectation value from the guiding wavefunction without GFMC.
		The inset zooms into the transition region $J_{xy}/J_z\in[-0.11,-0.10]$.
	}
	\label{fig:pyro_tau}
\end{figure}

We provide additional data on the convergence of the structure factor $S^{\pm}(\bm{q}=0)$ from \autoref{eq:Spm_q0}.
\autoref{fig:pyro_tau} shows how $S^{\pm}(\bm{q}=0)$ converges with the projection depth $\tau$ for $N=1024$.
We find that convergence is achieved reliably for $\tau\approx 10$ ``deep'' within the ferromagnet ($J_{xy}/J_z\leq-0.12$) and ``deep'' within the quantum spin liquid ($J_{xy}/J_z\geq-0.09$).
Hence, simulations were conducted for $\tau\approx 14$ in the ferromagnet and $\tau\approx 20$ in the quantum spin ice phase.
In contrast, the transition region $J_{xy}/J_z\in[-0.11,-0.10]$ requires a larger projection depth, as shown in the inset of \autoref{fig:pyro_tau}.
The order parameter $S^{\pm}(\bm{q}=0)$ decreases by approximately $2.5\%$ to $8\%$ between $\tau=20$ and $\tau=30$, and by $0.2\%$ to $2.7\%$ between $\tau=30$ and $\tau=40$.
To estimate the remaining error from the finite projection depth, we fit $S^{\pm}(\bm{q}=0)$ for $\tau\geq 10$ to $f(\tau) = c\,e^{-a\tau}+b$.
The extrapolated value $b$, which corresponds to the $\tau\to\infty$ limit, is shown using red markers in \autoref{fig:pyro_tau}.
It is visually indistinguishable from the data at the largest $\tau$.

We further show the order parameter computed with the guiding wavefunction alone (without GFMC) using black diamonds in \autoref{fig:pyro_tau}.
While the guide captures the overall trend of a small order parameter in the disordered phase and a strong signal in the ordered phase, it clearly struggles to obtain reliable data around the transition region, making a systematic finite-size scaling impossible.
In particular, the guide estimates vary irregularly between neighboring values of $J_{xy}/J_z$.
The variation suggests that optimization ended up in different local minima.
However, the GFMC algorithm reliably projects onto the true ground state, resulting in a smooth curve.

\section{Implemented operator classes}\label{app:hamiltonians}
All pre-implemented Hamiltonian classes provide the necessary functions and attributes introduced in \autoref{sec:pkg:hamiltonian}: \pyinline{M_total}, \pyinline{get_rates}, \pyinline{generate_state}, and \pyinline{diagonal_weight}.
The classes allow the couplings to be given as a uniform scalar or as per-edge/per-site/per-plaquette arrays.
Each constructor verifies that the resulting model is sign-problem-free.
The check can be bypassed with \pyinline{sign_check=False}, which is necessary when a Hamiltonian class is used as an off-diagonal operator.
All constructors accept a \pyinline{dtype} keyword (default \pyinline{jnp.float32}).
We recommend using \pyinline{jnp.float32}, as GPUs are substantially slower when handling \pyinline{jnp.float64}.

\paragraph{Spin.}
Spin-$S$ Hamiltonian with single- and two-body terms and a local Hilbert space of dimension $2S+1$ spanned by the values $-S, \dots, S$:
\begin{align}
	H \;=\; \sum_{\langle ij \rangle} \Big[ &J_{z,ij}\, S^z_i S^z_j
		+ J_{\pm,ij} \big( S^+_i S^-_j + S^-_i S^+_j \big)
		+ J_{\pm\pm,ij} \big( S^+_i S^+_j + S^-_i S^-_j \big) \Big] \notag \\
		&- \sum_i \Big[ h_{z,i}\, S^z_i + h_{\pm,i} \big( S^+_i + S^-_i \big) \Big],
	\label{eq:H_spin_app}
\end{align}
where $\langle ij \rangle$ runs over a given edge list.
The model is sign-problem-free for $J_{\pm}, \, J_{\pm\pm} \leq 0$ and $h_{\pm} \geq 0$.

\begin{lstlisting}
Spin.create(edges, S=0.5, J_z=0.0, J_pm=0.0, J_pmpm=0.0,
            h_z=None, h_pm=None, N=None, sign_check=True)
\end{lstlisting}

\paragraph{SpinHalf.}
This class is optimized for spin-1/2 Hamiltonians:
\begin{align}
	H \;=\; \sum_{\langle ij \rangle } \left(
		J_{x,ij}\, S^x_i S^x_j + J_{y,ij}\, S^y_i S^y_j + J_{z,ij}\, S^z_i S^z_j
	\right)
	\;-\; \sum_{i} \left( h_{z,i}\, S^z_i + h_{x,i}\, S^x_i \right)\,.
	\label{eq:H_spinhalf}
\end{align}
Internally, the transverse couplings are stored as $J_{\pm} = (J_x + J_y)/4$ (spin exchange) and $J_{\pm\pm} = (J_x - J_y)/4$ (double spin flip).
The model is sign-problem-free for $J_{\pm}, \, J_{\pm\pm} \leq 0$ and $h_x \geq 0$.
Notice that, when $J_{\pm\pm}>0$ and $h_x=0$, it is possible to swap $J_x$ and $J_y$.\\

\begin{lstlisting}
SpinHalf.create(edges, J_x=0, J_y=0, J_z=0, h_z=None, h_x=None,
                N=None, sign_check=True)
\end{lstlisting}

\paragraph{BoseHubbard.}
Bose--Hubbard model with hopping \pyinline{t} on each edge, a pairing amplitude \pyinline{eta} on each edge, on-site interaction \pyinline{U}, chemical potential \pyinline{mu}, an optional amplitude \pyinline{h_pm} multiplying both $b^\dagger_i$ and $b_i$, and a hard occupation cutoff \pyinline{n_max}:
\begin{align}
	H \;=\; -\sum_{\langle ij \rangle} \Big[ &t_{ij} \big( b^\dagger_i b_j + {\rm h.c.} \big)
		+ \eta_{ij} \big( b^\dagger_i b^\dagger_j + {\rm h.c.} \big) \Big] \notag \\
		&+ \sum_i \Big[ \tfrac{U_i}{2}\, n_i (n_i - 1) - \mu_i\, n_i
		- h_{\pm,i} \big( b^\dagger_i + b_i \big) \Big].
	\label{eq:H_bosehubbard_app}
\end{align}
Configurations are occupation numbers $0, \dots, n_{\rm max}$.
The model is sign-problem-free for $t, \eta, h_{\pm} \geq 0$.\\

\begin{lstlisting}
BoseHubbard.create(edges, t=0.0, eta=0.0, U=0.0, mu=0.0,
                   h_pm=None, N=None, n_max=4, sign_check=True)
\end{lstlisting}

\paragraph{ToricCode.}
Toric code on an $L_x \times L_y$ torus with periodic boundaries and stabilizer couplings \pyinline{A_s}, \pyinline{B_p} and local fields \pyinline{h_x}, \pyinline{h_z}~\cite{kitaev_fault_2003}.
\begin{align}
	H \;=\; -\sum_v A_s^{(v)} \prod_{i \in \star(v)} \sigma^x_i
		\;-\; \sum_p B_p^{(p)} \prod_{i \in \Box(p)} \sigma^z_i
		\;-\; \sum_i h_{z,i}\, \sigma^z_i \;-\; \sum_i h_{x,i}\, \sigma^x_i.
	\label{eq:H_toric_app}
\end{align}
Note that this class uses Pauli operators rather than spin-$1/2$ operators.
The spins live on the $2 L_x L_y$ edges of the square lattice.
The model is sign-problem-free for $A_s^{(v)}, h_{x,i} \geq 0$ everywhere.\\

\begin{lstlisting}
ToricCode.create(Lx, Ly, A_s=1.0, B_p=1.0, h_x=0.0, h_z=0.0, sign_check=True)
\end{lstlisting}

\paragraph{PlaquetteModel.}
Ring-exchange terms of amplitude \pyinline{g} on an arbitrary collection of plaquettes of equal length $\ell$, given as an $(N_{\rm plaquette}, \ell)$ array, together with a Rokhsar--Kivelson potential \pyinline{mu}~\cite{rokhsar_superconductivity_1988}:
\begin{align}
	H \;=\; -\sum_{p=(i,j,\ldots,k)\in \mathcal{P}} g_p \Big( S^+_i S^-_j \cdots S^-_k + {\rm h.c.} \Big)
	\;+\; \sum_{p\in\mathcal{P}} \mu_p\, n^{\rm flip}_p\,,
	\label{eq:H_plaquette_app}
\end{align}
where the sites in $p=(i,j,\ldots,k)$ are ordered around the loop, and $n^{\rm flip}_p$ indicates whether $p$ is flippable.
Note that $\ell$ must be even.
Writing the two flippable configurations as $\ket{\cw_p}$ and $\ket{\ccw_p}$, the same Hamiltonian reads:
\begin{align}
	H \;=\; -\sum_p g_p \Big( \ket{\ccw_p}\!\bra{\cw_p} + \ket{\cw_p}\!\bra{\ccw_p} \Big)
		\;+\; \sum_p \mu_p \Big( \ket{\cw_p}\!\bra{\cw_p} + \ket{\ccw_p}\!\bra{\ccw_p} \Big).
	\label{eq:H_plaquette_app_arrows}
\end{align}
The model is sign-problem-free for $g \geq 0$ with \pyinline{mu} of either sign.\\

\begin{lstlisting}
PlaquetteModel.create(plaquettes, g=None, mu=None, sign_check=True)
\end{lstlisting}

\section{User-defined operator classes}\label{app:user-defined}
We demonstrate how the user can construct their own operator classes using spin-$1/2$ degrees of freedom, which are encoded as $(-1,1)$.
\autoref{app:user-defined-hamiltonian} defines an $n$-body Hamiltonian that acts through a product of $\sigma^z$ and $\sigma^x$ on multiple sites.
\autoref{app:user-defined-observable} defines the diagonal $S^zS^z$ structure factor.

\subsection{An $n$-body Hamiltonian}\label{app:user-defined-hamiltonian}
The Hamiltonian consists of an on-site field and products of operators acting on given clusters of $n$ sites:
\begin{align}
	H = -J_z \sum_{c\,\in\, \mathcal{C}} \prod_{i\in c} \sigma^z_i
	    -J_x \sum_{c\,\in\, \mathcal{C}} \prod_{i\in c} \sigma^x_i
		-h \sum_i \sigma^z_i,\,
	\label{eq:user_nbody}
\end{align}
where $\mathcal{C}$ is a collection of $n$-site clusters.
Note that we use Pauli operators instead of spin operators.
The $\sigma^z$ part is diagonal and the $\sigma^x$ part flips every site of a cluster at once.
Therefore, each cluster contributes exactly one off-diagonal transition and $\Mtotal = \vert\mathcal{C}\vert + 1$.
The model is sign-problem-free for $J_x\geq 0$.\\

\begin{lstlisting}
import jax
import jax.numpy as jnp

class NBody:
    def __init__(self, clusters, J_z, J_x, h=0.0):
        self.clusters = jnp.asarray(clusters)          # (n_terms, n)
        self.J_z, self.J_x, self.h = J_z, J_x, h
        self.M_total  = 1 + self.clusters.shape[0]

    def get_rates(self, config, return_configs=False):
        diag  = self.diagonal_weight(config)
        off   = jnp.full(self.clusters.shape[0], -self.J_x)
        rates = jnp.concatenate([jnp.array([diag]), off])
        if not return_configs:
            return rates, None
        flips = jax.vmap(lambda c: config.at[c].multiply(-1))(self.clusters)
        return rates, jnp.concatenate([config[None, :], flips])

    def generate_state(self, config, idx):
        flipped = config.at[self.clusters[idx - 1]].multiply(-1)
        return jnp.where(idx == 0, config, flipped)

    def diagonal_weight(self, config):
        return (-self.J_z * jnp.sum(jnp.prod(config[self.clusters], axis=1))
                - self.h * jnp.sum(config))

jax.tree_util.register_pytree_node(
    NBody,
    lambda h: ((h.clusters,), (h.J_z, h.J_x, h.h)),
    lambda aux, children: NBody(children[0], *aux))
\end{lstlisting}
The class is registered as a \JAX{} pytree.

\subsection{$S^zS^z$ structure factor}\label{app:user-defined-observable}
The $S^zS^z$ structure factor at wavevector $\mathbf{k}$,
\begin{align}
	S^{zz}(\mathbf{k}) = \sum_{i<j} \left(e^{\mathrm{i} \mathbf{k}\cdot \mathbf{dr}_{ij}}
	           + e^{-\mathrm{i} \mathbf{k}\cdot \mathbf{dr}_{ij}}\right)S^z_iS^z_j
	           = \sum_{i<j} 2\cos\left( \mathbf{k}\cdot \mathbf{dr}_{ij}\right)S^z_iS^z_j\,,
	\label{eq:user_structure_factor}
\end{align}
is diagonal in the computational basis.
Hence, it only needs the \pyinline{diagonal_weight(config)} function.
$\mathbf{dr}_{ij}$ is the displacement between sites $i$ and $j$ and is passed to the constructor as one vector per pair so that the weights
$2\cos(\mathbf{k}\cdot\mathbf{dr}_{ij})$ can be precomputed.\\

\begin{lstlisting}
class PairSzSz:
    def __init__(self, pairs, dr, k):
        self.pairs = jnp.asarray(pairs)                # (n_pairs, 2)
        self.w = 2.0 * jnp.cos(jnp.asarray(dr) @ jnp.asarray(k))

    def diagonal_weight(self, config):
        s = config[self.pairs]
        return 0.25 * jnp.sum(self.w * s[:, 0] * s[:, 1])
\end{lstlisting}

\section{Implemented guiding wavefunctions}\label{app:guides}
We provide different Jastrow-type guides that can be directly used within different classes.
These wavefunctions have the form
\begin{align}
	\log \psi_T(x) =  \sum_{c\in\mathcal{C}} \Phi_c(c[x])\,,
	\label{eq:cluster_guide_app}
\end{align}
where the log-amplitude of each configuration is simply obtained by adding up the contributions of the finite-size clusters $c\in\mathcal{C}$.
This ansatz has two advantages.
First, it can easily be constructed to respect the symmetries of the system.
Second, it can be evaluated quickly, as a transition between configurations typically affects only a subset of the clusters.
The guides can be stored in and read from HDF5 via \pyinline{guide.save(path)} and \pyinline{Class.load(path)}, so that a previously trained guide can later be used for the Monte Carlo simulation.
All constructors accept a \pyinline{dtype} keyword (default \pyinline{jnp.float32}).

\paragraph{JastrowSpinHalf.}
The clusters are the spin pairs $c=\{i,j\}$ listed in \pyinline{pairs} together with the single sites,
\begin{align}
	\Phi_{\{i,j\}}(\sigma^z_i\, \sigma^z_j) \;=\; -\beta_{ij}\, \sigma^z_i\, \sigma^z_j\,,
	\qquad
	\Phi_{\{i\}}( \sigma^z_i) \;=\; h_i\, \sigma^z_i\,.
	\label{eq:phi_jastrow}
\end{align}
\pyinline{beta} is either a scalar, resulting in a uniform coupling, or an array whose length matches the number of pairs.
Similarly, \pyinline{h} can be a scalar or an array of size $N$.\\

\begin{lstlisting}
JastrowSpinHalf(N, pairs, beta, h=None)
\end{lstlisting}

\paragraph{JastrowSpin.}
Similar to \pyinline{JastrowSpinHalf()} but implemented for the \pyinline{Spin} class, with $x_i \in \{-S,\dots,S\}$:
\begin{align}
	\Phi_{\{i,j\}}(x_i, x_j) \;=\; -\beta_{ij}\, x_i\, x_j\,,
	\qquad
	\Phi_{\{i\}}(x_i) \;=\; h_i\, x_i - d_i\, x_i^2\,.
	\label{eq:phi_jastrowspin}
\end{align}
\begin{lstlisting}
JastrowSpin(N, pairs, beta, h=None, d=None)
\end{lstlisting}

\paragraph{JastrowBoseHubbard.}
Similar to \pyinline{JastrowSpinHalf()} but implemented for the \pyinline{BoseHubbard} class, with $x_i$ the occupation of site $i$:
\begin{align}
	\Phi_{\{i,j\}}(x_i, x_j) \;=\; -\beta_{ij}\, x_i\, x_j\,,
	\qquad
	\Phi_{\{i\}}(x_i) \;=\; h_i\, x_i - d_i\, x_i^2\,.
	\label{eq:phi_jastrowbosehubbard}
\end{align}
\begin{lstlisting}
JastrowBoseHubbard(N, pairs, beta, h=None, d=None)
\end{lstlisting}
\paragraph{ClusterCPS.}
The correlator product state~\cite{changlani_approximating_2009,mezzacapo_ground-state_2009} allows for different groups of clusters, where $N_G$ is the number of groups and the log-amplitude is obtained by:
\begin{align}
	\log\psi_T(x) \;=\; \sum_{g=1}^{N_G} \sum_{c\in\mathcal{C}_g} \Phi_g(c[x])\,.
	\label{eq:phi_clustercps}
\end{align}
$\mathcal{C}_g$ is a collection of clusters of equal size $L_g$ and $\Phi_g$ is a lookup table that maps a configuration to a real number.
The lookup table has $2^{L_g}$ entries since it is defined for the \pyinline{SpinHalf} class.

Each group $g$ is passed as \pyinline{[clusters, symmetries, parity]}, where \pyinline{clusters} is a two-dimensional array of shape $(|\mathcal{C}_g|, L_g)$ defining the clusters of the group. 
\pyinline{symmetries} contains all automorphisms to be applied to the lookup table, meaning that configurations that are connected by a given automorphism will have the same amplitude.
It is a two-dimensional array of size $(N_A,L_g)$, where $N_A$ is the number of automorphisms provided, or \pyinline{None} if no automorphism is imposed.
\pyinline{parity} is a Boolean (\pyinline{True} or \pyinline{False}) referring to the global spin flip $\sigma^z\to-\sigma^z$ symmetry.
The symmetry holds for $h_z=0$.
The spatial symmetries and parity reduce the table to its inequivalent orbits.
For example, the number of free parameters in the lookup table was reduced from $2^{18}=262144$ to $2110$ in the pyrochlore example of \autoref{sec:ex:pyro_nn}.\\

\begin{lstlisting}
ClusterCPS(groups, amplitudes=None, init='zero', seed=None)
\end{lstlisting}

\pyinline{groups} is the list of group specifications introduced above, with one entry per group.
An already trained set of lookup tables is passed as \pyinline{amplitudes} with one array per group.
If \pyinline{amplitudes=None}, the tables are initialized either to zeros (\pyinline{'zero'}) or to small Gaussian noise (\pyinline{'gaussian'}).
A seed can be set by \pyinline{seed}.

\paragraph{ClusterMLP.}
It has a similar structure to \pyinline{ClusterCPS}, but encodes each cluster group by a multilayer perceptron $\Phi_g$ instead of a lookup table.
See \texttt{NetKet}~\cite{SciPostPhysCodeb.7,SciPostPhysCodeb.7-r3.4} for a detailed description of such architectures.
All clusters of a group share one network, which may have several layers, so that the number of parameters is set by the width and depth of the network rather than by the cluster size $L_g$.
The arguments of the constructor are the direct analog of those of \pyinline{ClusterCPS}.
Each group $g$ is passed as \pyinline{[clusters, hidden]}.
\pyinline{clusters} is again a two-dimensional array of shape $(|\mathcal{C}_g|, L_g)$, and \pyinline{hidden} lists the widths of the hidden layers of that group's network.
For example, if \pyinline{hidden=[8, 4]} and the clusters of the group contain $L_g=6$ sites, the network will have the following dimensions: $(6, 8, 4, 1)$.
The input layer is fixed by the cluster size and the output layer is the final scalar.

\begin{lstlisting}
ClusterMLP(groups, params=None, seed=None)
\end{lstlisting}

The weights and biases $(W^{(l)}, b^{(l)})$ of all groups can be passed as \pyinline{params}, one pair per layer and group, and must be consistent with the layer dimensions implied by \pyinline{clusters} and \pyinline{hidden}.
If \pyinline{params=None}, the biases are set to zero and the weights are drawn from a Gaussian of width $0.01$.\\

In our benchmarks, we find that \pyinline{ClusterMLP} achieves lower energies than \pyinline{ClusterCPS}.
It is possible to train \pyinline{ClusterMLP} and load the amplitudes per state in \pyinline{ClusterCPS}.
However, \pyinline{ClusterMLP} breaks symmetries and potentially requires deeper projections within the GFMC algorithm.

\paragraph{ExactGroundState.}
We further provide the option to use the exact ground state as a guide, $\psi_T(x) = |\braket{x}{\Psi_0}|$, obtained by exact diagonalization of the given Hamiltonian.
This is mainly meant for validation.\\

\begin{lstlisting}
ExactGroundState(
    hamiltonian,
    log_amp=None,              # precomputed log|psi_0|; None diagonalizes hamiltonian 
    local_hilbert_space=None,  # None uses hamiltonian.local_hilbert_space
    N=None,                    # number of sites, defaults to hamiltonian.N
)
\end{lstlisting}

\section{User-defined guiding wavefunctions}\label{app:user-defined-guides}
We show how to implement a simple custom guide whose logarithmic amplitude for a configuration is given by the total magnetization times a control parameter $\beta$:
\begin{align}
	\log\psi_T(x) \;=\; \beta \sum_i \sigma^z_i\,.
	\label{eq:phi_custom_guide}
\end{align}
While the functions \pyinline{log_psi} and \pyinline{log_diff_from_configs} are independent of the underlying Hamiltonian, we also implement \pyinline{log_diff_from_operator}, which is tailored to the $n$-body Hamiltonian in \autoref{app:user-defined-hamiltonian}.
The local Hilbert space is encoded as $(-1,1)$, representing spin-$1/2$ degrees of freedom.

\begin{lstlisting}
class MagnetizationGuide:
    def __init__(self, beta):
        self.beta = jnp.asarray(beta)

    def log_psi(self, config, cache=None):
        # config is (N,) or (n, N)
        return self.beta * jnp.sum(config, axis=-1)

    def log_diff_from_configs(self, config, new_configs, cache=None):
        return self.log_psi(new_configs) - self.log_psi(config)

    def log_diff_from_operator(self, config, op, cache=None):
        # flipping the sites of cluster k changes the magnetization
        # by -2 sum_{i in cluster k} sigma_i; entry 0 is the diagonal
        dm = -2.0 * jnp.sum(config[op.clusters], axis=1)
        return self.beta * jnp.concatenate([jnp.zeros(1, dtype=dm.dtype), dm])

jax.tree_util.register_pytree_node(
    MagnetizationGuide,
    lambda g: ((g.beta,), None),
    lambda aux, children: MagnetizationGuide(*children),
)
\end{lstlisting}

\section{NetKet integration}\label{app:netket}
\PyGFMC{} offers a bridge for training \PyGFMC{} guiding wavefunctions directly with \texttt{NetKet}~\cite{SciPostPhysCodeb.7,SciPostPhysCodeb.7-r3.4}.
By passing the \PyGFMC{} Hamiltonian and the guide, together with the settings for the training procedure (see the \texttt{NetKet} documentation for more details), the variational optimization is carried out by \texttt{NetKet}.
The guide only requires \pyinline{log_psi} and a registration as a pytree whose floating-point children are the variational parameters (see the implementation of the guides in \autoref{app:guides} and \autoref{app:user-defined-guides} for details).
The Hamiltonian only requires \pyinline{get_rates}.
A new guide of the same class is returned with optimized parameters.

It is often desirable to constrain the variational states to respect the underlying symmetries of the Hamiltonian, such as particle number conservation.
For this purpose, one can pass \pyinline{initial_states} and a \pyinline{sampler} that restrict the states to a given symmetry sector.
The default option is \texttt{MetropolisExchange}, which conserves the particle number.\\

\begin{lstlisting}
from pygfmc.wavefunction import train_with_netket

guide = train_with_netket(
    hamiltonian,              # only get_rates is required
    guide,                    # only log_psi is required
    local_hilbert_space=None, # defaults to hamiltonian.local_hilbert_space
    N=None,                   # defaults to hamiltonian.N
    sampler=None,             # NetKet sampler; carries its own Hilbert space
    initial_states=None,      # (n, N) states seeding the Metropolis chains,
                              #   resampled to n_chains, so n is unrestricted
    n_iter=100,               # VMC iterations
    learning_rate=0.02,
    n_samples=1024,           # samples per iteration
    n_chains=16,              # parallel Metropolis chains
    n_discard_per_chain=8,    # warm-up samples dropped per chain
    sr=False,                 # stochastic reconfiguration
    diag_shift=1e-3,          # SR regularization
    qgt=None,                 # quantum geometric tensor implementation
    seed=0,
    show_progress=False,
    return_history=False,     # also return the per-iteration energies
)
\end{lstlisting}

\end{document}